\documentclass{llncs}
\usepackage{latexsym}
\begin{document}
\newcommand{\eat}[1]{}
\title{Termination analysis of logic programs using acceptability with general term orders}
\author{Alexander Serebrenik, Danny De Schreye}
\institute{Department of Computer Science, K.U. Leuven\\
Celestijnenlaan 200A, B-3001, Heverlee,
Belgium\\Email: \{Alexander.Serebrenik, Danny.DeSchreye\}@cs.kuleuven.ac.be}
\maketitle
\begin{center}
Technical report CW 291
\end{center}

\begin{abstract}
We present a new approach to termination analysis of logic programs. The
essence of the approach is that we make use of general term-orderings (instead
of level mappings), like it is done in transformational approaches to logic
program termination analysis, but that we apply these orderings directly to the
logic program and not to the term-rewrite system obtained through some 
transformation. We define some variants of acceptability, based on 
general term-orderings, and show how they are equivalent to LD-termination. 
We develop a demand driven, constraint-based approach to verify
these acceptability-variants.

The advantage of the approach over standard acceptability is that in some 
cases, where complex level mappings are needed, fairly simple term-orderings
may be easily generated. The advantage over transformational approaches is that
it avoids the transformation step all together.

\end{abstract}

\section{Introduction}
There are many different approaches to termination analysis of logic programs. One particular distinction is between {\it transformational\/} approaches and {\it ``direct''\/} ones. A transformational approach first transforms the logic 
program into an ``equivalent'' term-rewrite system (or, in some cases, into an equivalent functional program). Here, equivalence means that, at the very 
least, the termination of the term-rewrite system should imply the termination
of the logic program, for some predefined collection of queries\footnote{The approach of Arts~\cite{Arts:PhD} is exceptional in the sense that 
the termination of the logic program is concluded from a weaker property
of {\em single-redex normalisation\/} of the term-rewrite system.}. Direct approaches do not include such a transformation, but prove the termination directly
on the basis of the logic program. 

Besides the transformation step itself, there is one other technical difference
between these approaches. Direct approaches usually prove termination on the 
basis of a well-founded ordering over the natural numbers. More specifically,
they use a {\it level mapping}, which maps atoms to
natural numbers, and, they verify appropriate decreases of this level mapping 
on the atoms occuring in the clauses. On the other hand, transformational   
approaches make use of more general well-founded orderings over terms, such as 
reduction orders, or more specifically a simplification order, or others 
(see~\cite{Dershowitz:RTA}). 

At least for the direct approaches the systematic choice for level mappings
and norms, instead of general term orders, seems arbitrary and ad hoc. This
has been the main motivation for this paper. We present an initial study on 
the use of general well-founded term-orderings as a means of directly
proving the termination of logic programs---without intermediate transformation.
In particular, 
\begin{itemize}
\item
we study whether the theoretical results on acceptability can be reformulated on the
basis of general term orders,
\item
we evaluate to what extent the use of the general term orderings (instead
of level mappings) either improves or deteriorates the direct approaches.
\end{itemize}

To illustrate the latter point, consider the following program, that formulates
some of the rules for computing the repeated derivative of a linear function 
in one variable $u$ (see also~\cite{DM79:cacm}) :
\begin{example}
\label{example:rep:der}
\begin{eqnarray*}
&& \mbox{\sl d}(\mbox{\sl der}(u),1).\\
&& \mbox{\sl d}(\mbox{\sl der}(A),0) \leftarrow \mbox{\sl number}(A).\\
&& \mbox{\sl d}(\mbox{\sl der}(X+Y),DX+DY) \leftarrow \mbox{\sl d}(\mbox{\sl der}(X),DX), \mbox{\sl d}(\mbox{\sl der}(Y),DY).\\
&& \mbox{\sl d}(\mbox{\sl der}(X*Y),X*DY+Y*DX) \leftarrow \mbox{\sl d}(\mbox{\sl der}(X),DX), \mbox{\sl d}(\mbox{\sl der}(Y),DY).\\
&& \mbox{\sl d}(\mbox{\sl der}(\mbox{\sl der}(X)),DDX)\leftarrow \mbox{\sl d}(\mbox{\sl der}(X),DX), \mbox{\sl d}(\mbox{\sl der}(DX),DDX).
\end{eqnarray*}

Proving termination of this program on the basis of a level-mapping is hard. 
For this example, the required level-mapping is a non-linear function. 
In particular, a level mapping, such that:  
$\mid\!\!\mbox{\sl d}(X,Y)\!\!\mid = \|X\|,$ 
$\mid\!\!\mbox{\sl number}(X)\!\!\mid = 0,$ 
$\| \mbox{\sl der}(X) \| = 2^{\|X\|}, $ 
$\| X + Y \| = \mbox{\sl max}(\|X\|,\|Y\|) + 1,$ 
$\| X * Y \| = \mbox{\sl max}(\|X\|,\|Y\|) + 1,$ 
$\| u \| = 2,$ 
$\| n \| = 2,\;\;\mbox{\rm if $n$ is a number}$,%
would be needed. No automatic system for
proving termination on the basis of level mappings is able to generate such
mappings. Moreover, we believe, that it would be very difficult to extend 
existing systems to support generation of appropriate non-linear mappings.
$\hfill\Box$\end{example}

Although we have not yet presented our general-well-founded term ordering 
approach, it should be intuitively clear, that we can capture the decrease in 
order between the $\mbox{\sl der}(X)$ and $DX$  by using an 
ordering on terms that gives the highest ``priority'' to the functor 
{\sl der}. 

As an example of the fact that moving to general ordering can also introduce
deterioration, consider the following program from~\cite{DeSchreye:Decorte:NeverEndingStory,Decorte:DeSchreye:Vandecasteele}.
\begin{example}
\begin{eqnarray*}
&& \mbox{\sl conf}(X)\leftarrow \mbox{\sl delete$_2$}(X,Z), \mbox{\sl delete}(U,Y,Z), \mbox{\sl conf}(Y).\\
&& \mbox{\sl delete$_2$}(X,Y)\leftarrow \mbox{\sl delete}(U,X,Z), \mbox{\sl delete}(V,Z,Y).\\
&& \mbox{\sl delete}(X,[X|T],T).\\
&& \mbox{\sl delete}(X,[H|T],[H|T1])\leftarrow \mbox{\sl delete}(X,T,T1). 
\end{eqnarray*}

Note that by reasoning in terms of sizes of terms, we can infer that the size 
decreases by 2 after the call to $\mbox{\sl delete}_2$ predicate in the
first clause and then increases by 1 in the subsequent call to the 
{\sl delete} predicate. In total, sizes allow to conclude a decrease. 
Reasoning in terms 
of order relations only, however, does not allow to conclude the overall 
decrease from the inequalities $\mbox{\sl arg3} < \mbox{\sl arg2}$ for the 
{\sl delete} predicate and $\mbox{\sl arg1} > \mbox{\sl arg2}$
for the $\mbox{\sl delete}_2$ predicate.
$\hfill\Box$\end{example}

As can be expected, theoretically both approaches are essentially equivalent.
We will introduce a variant of the notion of acceptability, based on general
term orders, which is again equivalent to termination in a similar way as 
in the level mapping based approach. On the more practical level, as illustrated
in the two examples above, neither of the approaches is strictly better: the
general term orders provide a larger set of orders to select from (in 
particular, note that orders based on level mappings and norms are a term 
order), the level mapping approach provides arithmetic, on top of mere 
ordering. 

In the remainder of this paper, we will start off from a variant of the notion of 
{\it acceptability with respect to a set}, as introduced in \cite{DeSchreye:Verschaetse:Bruynooghe},
obtained by replacing level mappings by term orderings. We show how 
this variant of acceptability
remains equivalent to termination under the left-to-right selection rule, for certain goals. 
Then, we illustrate how this result can be used to prove termination with some examples. We also provide a variant of the {\em acceptability} condition, as 
introduced in~\cite{Apt:Pedreschi:Studies}, and discuss advantages and 
disadvantages of each approach. 
Next, we discuss automation of the approach. We elaborate on a demand-driven method to set-up 
and verify sufficient preconditions for termination. In this method, the aim is to 
derive---in, as much as possible, a constructive way---a well-founded ordering over the set 
of all atoms and terms of the language underlying the program, that satisfies the termination 
condition.


\section{Preliminaries}
\subsection{Term ordering}
An {\em order} over a set $S$ is an irreflexive, asymmetric and transitive
relation $>$ defined on elements of $S$.  As usual, $s\geq t$ denotes that
either $s > t$ or $s =_> t$. $s=_>t$ denotes that $s$ and $t$ are equal under
the order, but not necessary identical, and $s\| t$ denotes that $s$ and $t$ 
are incomparable. An ordered set $S$ is said to be {\em well-founded}
if there are no infinite descending sequences $s_1 > s_2 > \ldots$ of elements
of $S$. If the set $S$ is clear from the context we will say that the order,
defined on it, is well-founded.

\begin{definition}
Let $>$ be an order on a set $T$. An order $\succ$ defined on a set 
$S\supseteq T$ is called an {\em extension of $>$\/} if for any 
$t_1, t_2\in T\;\;t_1 > t_2$ implies $t_1\succ t_2$.
\end{definition}

The study of termination of term-rewriting systems caused intensive study of
term orderings. A number of useful properties of term orderings were 
established.

\begin{definition}
Let $>$ be an ordering on terms.
\begin{itemize}
\item If $s_1 > s_2$ implies $f({\bar t_1},s_1,{\bar t_2}) > f({\bar t_1},s_2,{\bar t_2})$ and $p({\bar t_1},s_1,{\bar t_2}) > p({\bar t_1},s_2,{\bar t_2})$ 
for any sequences of terms ${\bar t_1}$ and ${\bar t_2}$, function symbol $f$ 
and predicate $p$, then $>$ is called {\em monotone}.
\item If for any term $f({\bar t_1},s,{\bar t_2})$ holds that  $f({\bar t_1},s,{\bar t_2}) > s$, then $>$ is said to have the {\em subterm property}.
\end{itemize}
\end{definition}

The following are examples of order relations: $>$ on the set of numbers,
lexicographic order on the set of strings (this is a way the entries are 
ordered in dictionaries), multiset ordering and recursive path 
ordering~\cite{Dershowitz:RTA}. 

For our purposes monotonicity and subterm properties are too restrictive.
Thus, we assign to each predicate or functor a subset of argument positions,
such that for the argument positions in this subset the specified properties
hold. We will say that a predicate $p$ (a functor $f$) is monotone (has a 
subterm property) on a specified subset of argument positions. The formal
study of these weaker notions may be found in Section~\ref{section:weaker:notions}.

\begin{example}
Let $f$ be a functor or arity two, and $a$, $b$ two terms, such that 
$a > b$. Let $f$ be monotone in the first argument position. Then,
$f(a,c) > f(b,c)$ holds for any term $c$, but there might be some term
$c$, such that $f(c,a) \not > f(c,b)$.
\end{example}

\subsection{Logic Programs}

We follow the standard notation for terms and atoms. A {\em query} is a 
finite sequence of atoms. Given an atom $A$, $\mbox{\sl rel}(A)$ denotes
the 
predicate occuring in $A$. $Term_P$ and $Atom_P$ 
denote, respectively, sets of all terms and atoms that can be constructed from
the language underlying $P$. The extended Herbrand Universe $U^E_P$ (the 
extended Herbrand base $B^E_P$) is a quotient set of $Term_P$ ($Atom_P$) modulo
the variant relation. 

We refer to an SLD-tree constructed using the left-to-right selection rule of
Prolog, as an LD-tree. We will say that a goal $G$ {\it LD-terminates} for
a program $P$, if the LD-tree for $(P,G)$ is finite.

The following definition is borrowed from~\cite{Apt:Book}.
\begin{definition}
Let $P$ be a program and $p$, $q$ be predicates occuring in it.
\begin{itemize}
\item We say that {\em $p$ refers to $q$ in $P$\/} if there is a clause in $P$
that uses $p$ in its head and $q$ in its body.
\item We say that {\em $p$ depends to $q$ in $P$\/} and write $p\sqsupseteq q$,
if $(p,q)$ is in the transitive, reflexive closure of the relation {\em refers to}.
\item We say that {\em $p$ and $q$  are mutually recursive\/} and write $p\simeq q$, if $p\sqsupseteq q$ and $q\sqsupseteq p$.
\end{itemize}
\end{definition}

If $p\sqsupseteq q$, but $q\not \sqsupseteq p$ we also write $p\sqsupset q$.
\section{Revising properties of orders}
\label{section:weaker:notions}
Unfortunately, monotonicity and subterm properties even being common for
the orders, useful for the termination proofs do not hold for orders like
list-size norm based order. Indeed, $\|[[1,2,3],4]\| = 2 < 3 = \|[1,2,3]\|$. 
Moreover, as we will see if those properties are assumed the notion of rigidity
deteriorates to groundedness.

Thus, we extend the notions above further, to incorporate these orders
as well. We start with some preliminary remarks.

\subsection{Preliminaries}

We start by recalling the classical definition of the characteristic
function of a set.

\begin{definition}
Let $I$ be a set.
The {\em characteristic function $\chi_I$} is defined as following:
\[
\chi_I(x) = \left\{
\begin{array}{ll}
1 &\;\;\; x\in I \\
0 &\;\;\; x\not\in I
\end{array}
\right.
\]
\end{definition}

We associate with each functor $f/n$ some set 
$I_{f/n}\subseteq \{1,\ldots,n\}$. Then, for term $s$ we can extend
the definition above in the following way:
\begin{definition}
Let $L$ be a language.
Let ${\cal I}$ be the family of sets, associated with functors, such that
$I_{f\!/\!n}\in {\cal I}$ exists for every $f/n\in L$.
Let $s$ be a term and $\overrightarrow{v}$ a vector of integers.
Then the {\em characteristic function $\chi_s$} is defined as following:
\[
\chi^{\cal I}_s(\overrightarrow{v}) = \left\{
\begin{array}{ll}
1 & \overrightarrow{v}\;\mbox{\rm is an empty vector}\\
1 & s\;\mbox{\rm is a variable or a constant}\\
\chi_{I_{f\!/\!n}}(i_1) * \chi^{\cal I}_{t_{i_1}}(\overrightarrow{i_2,\ldots,i_k}) &\;\;\; \mbox{\rm if}\;\; s = f(t_1,\ldots,t_n) \\
\end{array}
\right.
\]
\end{definition}

We illustrate this definition by the following example.
\begin{example}
\label{example:branches}
Let $s$ be $f(a, g(b, h(c, X)))$ and let $I_{f/2}$ be $\{1,2\}$,
$I_{g/2}$ be $\{2\}$ and $I_{h/2}$ be $\{1\}$. Then, the following holds
$\chi^{\cal I}_s(\overrightarrow{1}) = 1$, 
$\chi^{\cal I}_s(\overrightarrow{2, 1}) = 0$,
$\chi^{\cal I}_s(\overrightarrow{2, 2}) = 1$,
$\chi^{\cal I}_s(\overrightarrow{2, 2, 1}) = 1$, 
$\chi^{\cal I}_s(\overrightarrow{2, 2, 2}) = 0$.
\end{example}

The definition above also suggests that the vector notation denotes
a branch in the tree representation of term. When no confusion can be caused
we will also talk about the value of the characteristic function for the
subterm, that is denoted by a vector and not for the vector itself. 
\begin{example}
Continuing Example~\ref{example:branches} we can rewrite the values
of the characteristic function as following:
$\chi^{\cal I}_s(a) = 1$, $\chi^{\cal I}_s(b) = 0$,
$\chi^{\cal I}_s(h(c,d)) = 1$,
$\chi^{\cal I}_s(c) = 1$, $\chi^{\cal I}_s(X) = 0$.
\end{example}

\begin{proposition}
Let $s$ be a term and $t$ be a subterm of it. Then, if 
$\chi^{\cal I}_s(t) = 0$,
for any subterm $t'$ of $t$ holds that $\chi^{\cal I}_s(t') = 0$.
\end{proposition}
\begin{proof}
Immediately, from the fact that the vector defining $t$ is a prefix of the
vector defining $t'$. 
\end{proof}

\subsection{Monotonicity - revised}

As we've mentioned already the properties of monotonicity and subterm
might be seen as very restrictive. Thus, we introduce the following
properties, that relax those notions.

\begin{definition}
Let $S$ be a set of terms over the language $L$, and let $>$ be an order 
defined on $S$. We say that $>$ is {\em partially monotone} if 
for every functor $f/n$ in $L$ exists 
${\cal M}_{f/n}\subseteq \{1,\ldots,n\}$, such that if $t_1 > t_2$
and $i\in {\cal M}_{f/n}$ then for any sequences of terms $\bar{s}$,
(occupying positions $1\ldots i-1$) and $\bar{u}$ (occupying positions
$i+1\ldots n$) holds
$f(\bar{s},t_1,\bar{u}) > f(\bar{s},t_2,\bar{u})$.
The sets ${\cal M}_{f/n}$ are called {\em defined} by $>$.
\end{definition}

\begin{example}
The order defined by the list-size norm is not monotone, but is partially
monotone. It defines ${\cal M}_{./2} = \{2\}$ and for other functors $f/n$
holds that ${\cal M}_{f/n} = \emptyset$.
\end{example}

\begin{proposition}
Let $S$ be a set. If $>$, defined on $S$ is monotone, than $>$ is partially
monotone.
\end{proposition}
\begin{proof}
Immediate from the definitions. ${\cal M}_{f/n} = \{1,\ldots,n\}$.
\end{proof}

We denote the collection of all ${\cal M}_{f/n}$ by ${\cal M}$.

\begin{lemma}
\label{mon:lemma}
Let $s$ be a term, and $X_{(i)}$ be a variable occurrence in it,
such that $\chi^{\cal M}_s(X_{(i)}) = 1$.
Then, $t_1 > t_2$ implies 
$s\{t_1\rightarrow X_{(i)}\} > s\{t_2\rightarrow X_{(i)}\}$.
\end{lemma}
\begin{proof}
Let $\overrightarrow{i_1,\ldots,i_k}$ be a vector that denotes $X_{(i)}$.
Then, by definition of the characteristic function 
$\chi_{{\cal M}_{f_1/n_1}}(i_1) * \ldots * \chi_{{\cal M}_{f_k/n_k}}(i_k) = 1$.
That is for any $j$ holds $\chi_{{\cal M}_{f_j/n_j}}(i_j) = 1$.

Since $\chi_{{\cal M}_{f_k/n_k}}(i_k) = 1$, $i_k\in {\cal M}_{f_k/n_k}$.
Thus, $t_1 > t_2$ implies $f(\bar{u_1},t_1,\bar{u_2}) > 
f(\bar{u_1},t_2,\bar{u_2})$ for any sequences $\bar{u_1}$ and
$\bar{u_2}$. In particular, this holds for the sequences appearing in $s$.
Repetitive application of this argument proves the lemma.
\end{proof}

The following example shows, that if $\chi^{\cal M}_s(X_{(i)}) = 0$
the lemma does not necessary hold.
\begin{example}
Let $>$ be order based on the list-size norm, and let $s$ be $[X,5]$.
The only variable in $s$ is $X$ and $\chi^{\cal M}_s(X) = 0$.
On the other hand, let $t_1$ be $[1,2]$ and $t_2$ be $[3]$. According
to the definition of $>$, $t_1 > t_2$ holds. However,
$s\{t_1\rightarrow X\} = [[1,2],5]$, 
$s\{t_2\rightarrow X\} = [[2],5]$ and $[[1,2],5] =_> [[2],5]$.
\end{example}

\subsection{The subterm property - revised}.

Now we are going to discuss the extension of notion of the subterm property.

\begin{definition}
Let $S$ be a set of terms over the language $L$, and let $>$ be an order 
defined on $S$. We say that $>$ has a {\em partially subterm property} if 
for every functor $f/n$ in $L$ exists 
${\cal S}_{f/n}\subseteq \{1,\ldots,n\}$, such that if 
$i\in {\cal S}_{f/n}$ then for any sequences of terms $\bar{s}$,
(occupying positions $1\ldots i-1$) and $\bar{u}$ (occupying positions
$i+1\ldots n$) holds
$f(\bar{s},t,\bar{u}) > t$.
The sets ${\cal S}_{f/n}$ are called {\em defined} by $>$.
\end{definition}

\begin{example}
The order defined by the list-size norm does not have a subterm property,
but it has a partial subterm property.
It defines ${\cal S}_{./2} = \{2\}$ and for other functors $f/n$
holds that ${\cal S}_{f/n} = \emptyset$.
\end{example}

\begin{proposition}
Let $S$ be a set of atoms or terms. If $>$, defined on $S$ has a subterm 
property, than $>$ has a partially subterm property.
\end{proposition}
\begin{proof}
Immediate from the definitions. ${\cal S}_{f/n} = \{1,\ldots,n\}$.
\end{proof}

We denote the collection of all ${\cal S}_{f/n}$ by ${\cal S}$.

\begin{lemma}
\label{sub:lemma}
Let $s$ be a term, and $t$ be a subterm occuring in it,
such that $\chi^{\cal S}_s(t) = 1$.
Then, $s > t$.
\end{lemma}
\begin{proof}
Let $\overrightarrow{i_1,\ldots,i_k}$ be a vector that denotes $t$.
Then, by definition of the characteristic function 
$\chi_{{\cal S}_{f_1/n_1}}(i_1) * \ldots * \chi_{{\cal S}_{f_k/n_k}}(i_k) = 1$.
That is for any $j$ holds $\chi_{{\cal S}_{f_j/n_j}}(i_j) = 1$.

Since $\chi_{{\cal S}_{f_k/n_k}}(i_k) = 1$, $i_k\in {\cal S}_{f_k/n_k}$.
Thus, $f(\bar{u_1},t,\bar{u_2}) > t$
for any sequences $\bar{u_1}$ and $\bar{u_2}$. In particular, this holds for 
the sequences appearing in $s$.
Repetitive application of this argument and the transitivity of $>$ 
proves the lemma.
\end{proof}

The following example shows, that if $\chi^{\cal S}_s(t) = 0$
the lemma does not necessary hold.
\begin{example}
Let $>$ be order based on the list-size norm, and let $s$ be $[[1,2,3],5]$.
Observe, that $\chi^{\cal S}_s([1,2,3]) = 0$.
On the other hand, $[[1,2,3],5] < [1,2,3]$.
\end{example}

\section{Term-acceptability with respect to a set}
In this section we present and discuss some of the theory we developed
to extend acceptability to general term orders. In the literature, there are 
different variants of acceptability. The most well-known of these is the 
acceptability as introduced by Apt and Pedreschi~\cite{Apt:Pedreschi:Studies}.
This version is defined and verified on the level of ground instances of
clauses, but draws its practical power mostly from the fact that termination is
proved for {\em any bounded\/} goal. Here, boundedness is a notion related to
the selected level mapping and requires that the set $\{|G\theta|\;\mid\;\theta
\;\mbox{\rm is a grounding substitution for goal}\;G\}$ is bounded in the 
natural numbers, where $|\cdot|: B_P\rightarrow {\cal N}$ denotes the level 
mapping.

Another notion of acceptability is the ``acceptability with respect to a set 
of goals'', introduced by De Schreye et. al.\ in~\cite{DeSchreye:Verschaetse:Bruynooghe}. This notion allows to prove termination with respect to any set of
goals of interest. However, it relies on procedural concepts, such as calls
and computed answer substitution. It was designed to be verified through
global analysis, for instance through abstract interpretation.

A variant of acceptability w.r.t. a set that avoids the drawbacks of using 
procedural notions and that can be verified on a local level was designed
in~\cite{Decorte:DeSchreye:Vandecasteele}. This variant required that the 
goals of interest are {\em rigid\/} under the given level mapping. Here, 
rigidity means that $|G\theta| = |G|$, for any substitution $\theta$, where
$|\cdot|: B^E_P\rightarrow {\cal N}$ now denotes a generalised level mapping,
defined on the extended Herbrand base.

Comparing the notions of boundedness and rigidity in the context of a level 
mapping based approach, it is clear that boundedness is more general than 
rigidity. If the level mapping of a goal is invariant under substitution, then the level mapping is bounded on the set of instances of the goal, but not 
conversely.

Given the latter observation and given that acceptability of~\cite{Apt:Pedreschi:Studies} is a more generally known and accepted notion, we started our work by
generalising this variant.

However, it turned out that generalising the concept of boundedness to general
term orders proved to be very difficult. We postpone the discussion on this 
issue until after we formulated the results, but because of these complications, we only arrived at generalised acceptability conditions that are useful
in the context of well-mode, simply-mode programs and goals.

Because of this, we then turned our attention to acceptability
with respect to a set. Here, the generalisation of rigidity was less 
complicated, so that in the end we obtained the strongest results for this
variant of acceptability. Therefore, we first present term-acceptability with
respect to a set of goals. We need the following notion.

\begin{definition}~\cite{Decorte:DeSchreye:98} 
Let $P$ be a definite program and $S$ be a set of atomic queries. 
The {\em call set}, $\mbox{\sl Call}(P,S)$, is the set of all atoms $A$,
such that a variant of $A$ is a selected atom in some derivation for
$P\cup \{\leftarrow Q\}$, for some $Q\in S$ and under the left-to-right 
selection rule.
\end{definition}

To illustrate this definition recall the following 
example~\cite{Apt:Book,Decorte:DeSchreye:Vandecasteele}.

\begin{example}
\label{example:permute}
\begin{eqnarray*}
&& {\mbox {\sl permute}}([],[]).\\
&& {\mbox {\sl permute}}(L,[El|T])\leftarrow {\mbox {\sl delete}}(El,L,L1), {\mbox {\sl permute}}(L1,T).\\
&& {\mbox {\sl delete}}(X,[X|T],T).\\
&& {\mbox {\sl delete}}(X,[H|T],[H|T1])\leftarrow {\mbox {\sl delete}}(X,T,T1).
\end{eqnarray*}
Let $S$ be $\{{\mbox {\sl permute}}(t_1, t_2)|\;t_1\;\mbox{\rm is a nil-terminated list and}\;t_2\;\mbox{\rm is a free variable}\}$. 
Then, $\mbox{\sl Call}(P,S) =$
\[S \cup \{{\mbox{\sl delete}}(t_1,t_2,t_3)|\;t_1, t_3\;\;\mbox{\rm are free variables and}\;t_2\;\mbox{\rm is a nil-terminated list}\}.\] Such information
about $S$ could for instance be expressed in terms of the rigid types of~\cite{Janssens:Bruynooghe} and $\mbox{\sl Call}(P,S)$ could be computed using the 
type inference of~\cite{Janssens:Bruynooghe}. 
$\hfill\Box$\end{example}

The following definition generalises the notion of acceptability w.r.t. 
a set~\cite{Decorte:DeSchreye:98} in two ways: 1) it generalises it to general
term orders, 2) it generalises it to mutual recursion, using
 the standard notation of mutual recursion~\cite{Apt:Book}.
\begin{definition}
\label{def:taset}
Let $S$ be a set of atomic queries and $P$ a definite 
program. $P$ is {\em term-acceptable w.r.t.\ $S$} if there exists a 
well-founded order $>$, such that
\begin{itemize}
\item for any $A\in\mbox{\sl Call}(P,S)$ 
\item for any clause $A'\leftarrow B_1,\ldots,B_n$ in $P$, such that 
$\mbox{\rm mgu}(A,A') = \theta$ exists,
\item for any atom $B_i$, such that $\mbox{\sl rel}(B_i)\simeq \mbox{\sl rel}(A)$
\item for any computed answer substitution $\sigma$ for 
$\leftarrow (B_1, \ldots, B_{i-1})\theta$:
\[A > B_i\theta\sigma\]
\end{itemize}
\end{definition}

In order to establish the connection between term-acceptability w.r.t.
a set $S$ and  LD-termination for queries in $S$ we recall the notion of
{\em directed sequence\/} and related results, introduced by Verschaetse
in~\cite{Verschaetse:thesis}.

\begin{definition}
Let $G_0, G_1, G_2,\ldots,\theta_1, \theta_2,\ldots$ be a derivation with 
selected atoms $A_0, A_1, A_2,\ldots$ and applied renamed clauses
$H^i\leftarrow B^i_1,\ldots, B^i_{n_i}$ ($i = 1,2,\ldots$). We say that $A_k$ 
is a {\em direct descendant\/} of $A_i$, if $k > i$  and $A_k$ is the atom
$B^{i+1}_j\theta_{i+1}\ldots\theta_k$, ($1\leq j\leq n_{i+1}$).
\end{definition}

\begin{definition}
Let $G_0, G_1, G_2,\ldots,\theta_1, \theta_2,\ldots$ be a derivation with 
selected atoms $A_0, A_1, A_2,\ldots$. A {\em subsequence\/} of derivation 
steps, 
$G_{i(0)}, G_{i(1)}, \ldots, \theta_{i(0)+1}, \ldots$ with selected atoms 
$A_{i(0)}, A_{i(1)}, A_{i(2)},\ldots$ is 
{\em directed}, if for each $k$ ($k > 1$), $A_{i(k)}$ is a direct descendant
of $A_{i(k-1)}$ in the given derivation.
\end{definition}

\begin{definition}
A derivation $G_0, G_1, G_2,\ldots,\theta_1, \theta_2,\ldots$ is {\em directed\/} if it is its own directed subsequence.
\end{definition}

Verschaetse~\cite{Verschaetse:thesis} proved also the following lemma:
\begin{lemma}
\label{Verschaetse:lemma}
Let $P$ be a definite program and $A$ an atom. If $(P,\leftarrow A)$ has an 
infinite derivation, then it has an infinite directed derivation.
\end{lemma}

Based on this results we prove the characterisation of the LD-termination
in terms of term-acceptability w.r.t. a set.

\begin{theorem}
\label{taset:term}
Let $P$ be a program. $P$ is term-acceptable w.r.t.\ a set of atomic
queries $S$ if and only if $P$ is LD-terminating for all queries in $S$.
\end{theorem}
\begin{proof}
\begin{itemize}
\item[$\Rightarrow$]
Let $A\in S$, and assume that $P\cup \{\leftarrow A\}$ has an infinite 
LD-derivation.
This derivation contains an infinite directed subsequence, that is a 
subsequence of goals $G_{i(0)}, G_i(1), \ldots$ such that the selected atom of 
$G_{i(k+1)}$, $A_{k+1}$, is a direct descendant of the selected atom of 
$G_{i(k)}$, $A_k$. There is some $k_0$, such that for any $m,n > k_0$
holds $\mbox{\sl rel}(A_m)\simeq \mbox{\sl rel}(A_n)$. Let $k$ be greater
than $k_0$.

Then there is a clause $H\leftarrow B_1,\ldots,B_n$,
such that the $\mbox{\rm mgu}(A_k, H) = \theta$ exists and for some
$j$ holds that $A_{k+1} = B_j\theta\sigma$, where $\sigma$ is a computed
answer substitution for $\leftarrow B_1,\ldots,B_{j-1}$. Observe that the 
choice of $k$ implies that $\mbox{\sl rel}(B_j)\simeq \mbox{\sl rel}(H)$.

Since $A_k$ is one of the selected atoms in $\mbox{\sl Call}(P,S)$ 
the condition of the term-acceptability w.r.t. $S$ is applicable. Thus, 
$A_k > B_j\theta\sigma$, i.e., $A_k > A_{k+1}$. By proceeding in this
way we construct an infinitely decreasing chain of atoms, contradicting
the well-foundedness of $>$.
\item[$\Leftarrow$]
Let $G$ be in $\mbox{\sl Call}(P,S)$, and let be $H\leftarrow B_1,\ldots,B_n$
be a clause in $P$, such that the $\mbox{\rm mgu}(H,G) = \theta$ exists.
We define a relation $\succ$, such that 
$G \succ B_i\theta\sigma$ for any $1\leq i\leq n$, where
$\sigma$ is the computed answer substitution for $\leftarrow (B_1,\ldots,B_{i-1})\theta$. Let $>$ be the transitive closure of $\succ$.

We start from the following observation. If $A\succ B$ there
is a derivation, started by $A$, and having a goal with selected atom $B$
in it. Similarly, if $A\succ B$ and $B\succ C$ there
is a derivation, started by $A$, and having a goal with selected atom $B$
in it, followed (not necessary immediately) by a goal with the selected atom
$C$. We extend this observation to $>$, and claim that if $A > B$ 
then there is a derivation for $A$, having a goal with the selected atom $B$
in it.

An additional observation is that if $X > Y$ is defined for some $X$ and $Y$, 
then $X$ and $Y$ are are in $\mbox{\sl Call}(P,S)$.

In particular, if $A> A$ there is a derivation for $A$ that has a goal $G$ 
with the selected atom $A$ in it. Thus, we can continue the derivation
from $G$ in the same way we have constructed a derivation from $A$ to $G$. 
This process can be done forever, contradicting that $P$ terminates for $A$. 

If $A > B$, $B > C$ from the observation above we construct a derivation from
$A$ that has a goal with the selected atom $B$, and then continue by mimicking 
the derivation from $B$, that has a goal with the selected goal $C$. Thus,
$A > C$. This also proves the asymmetry (if $A > B$ and $B > A$, then $A > A$,
by transitivity, and this causes contradiction by irreflexivity).  

The well-foundedness follows from the finiteness of all the derivations.
The term-acceptabity w.r.t. $S$ follows immediately from the definition of $>$.
\end{itemize}
\end{proof}

We postpone applying the Theorem~\ref{taset:term} to
Example~\ref{example:permute} until a more syntactic way of
verifying term-acceptability w.r.t. a set is developed.

To do this, we extend the 
sufficient condition of~\cite{DeSchreye:Verschaetse:Bruynooghe}, that impose
the additional requirement of rigidity of the level mapping on the call set,
to the case of general term orders and study the notion of {\em rigidity}.

\section{Rigidity}
First we adapt the notion of rigidity to general term orders.
\begin{definition}(see also~\cite{Bossi:Cocco:Fabris})
The term or atom $A\in U^E_P \cup B^E_P$ is called {\em rigid} w.r.t. $>$ 
if for any substitution $\theta$, $A =_> A\theta$. In this case $>$ is said
to be {\em rigid on} $A$.
\end{definition}

The notion of the rigidity on a term (an atom) is naturally extended to 
the notion of rigidity on a set of atoms (terms). In particular, we will be 
interested in orders that are rigid on $\mbox{\sl Call}(P,S)$ for some $P$ 
and $S$. Intuitively, the rigid order ignores argument positions that might be
occupied by free variables in $\mbox{\sl Call}(P,S)$.

\begin{proposition}
The following simple properties of rigid terms and atoms hold:
\begin{enumerate}
\item If $A$ is ground, then $A$ is rigid.
\item If $A$ is rigid w.r.t. $>$, then $A\theta$ is rigid w.r.t. $>$ for
any substitution $\theta$.
\end{enumerate}
\end{proposition}

Also, observe, that similarly to the notion of rigidity w.r.t. a norm,
$A =_> B$ and $A$ is rigid does not imply that $B$ is rigid.

Let $>$ be an order and $s$ be a term, that is not rigid. Then there must
occur some variables in $s$ whose substitution causes the disequality to hold.
We want to identify each of their occurrences.

\begin{definition}(see~\cite{Bossi:Cocco:Fabris})
Let $>$ be an order and $s$ be a term. The $i$th occurrence $X_{(i)}$,
of a variable $X$ in the term $s$ is {\em relevant w.r.t. $>$} whenever there
exists a replacement $\{t\rightarrow X_{(i)}\}$ of the term $t$ for the
$i$th occurrence of $X$ in $s$, such that $s\{t\rightarrow X_{(i)}\}\not =_> s$.
$\mbox{\rm VREL}_>(s)$ is the set of all the relevant occurrences of variables 
in $s$.
\end{definition}

Bossi, Cocco and Fabris~\cite{Bossi:Cocco:Fabris} proved that for the 
semi-linear norm $\|\cdot\|$ the term $s$ is rigid w.r.t.\ $\|\cdot\|$
if and only if $\mbox{\rm VREL}_{\|\cdot\|}(s) = \emptyset$. This allowed to
reduce a check of the rigidity w.r.t. the semi-linear norm to a syntactical 
condition. As the following example demonstrates, the assumption of 
semi-linearity is essential.
\begin{example}(\cite{Bossi:Cocco:Fabris})
Let $\| \|_{\mbox{\rm bal}}$ be the following norm:
\[
\|t\|_{\mbox{\rm bal}} = \left\{
\begin{array}{ll}
0 & \;\;t\;\mbox{\rm is void or is not a tree}\\
0 & \;\;t\;\mbox{\rm is a ${\mbox{\sl tree}}(a,l,r)$ and $l = r$}\\
1 & \;\;t\;\mbox{\rm is a ${\mbox{\sl tree}}(a,l,r)$ and $l\not = r$},
\end{array}
\right.
\]
where $=$, as usual, means syntactical equality.

This norm is not semi-linear, and indeed the term ${\mbox{\sl tree}}(a,X,X)$
is rigid w.r.t. it since for any substitution of the variable $X$, the two
subtrees remain syntactically equal. Nevertheless, $\mbox{\rm VREL}({\mbox{\sl tree}}(a,X,X)) = \{X_{(1)},X_{(2)}\}$ since by replacing one single
occurrence of the variable $X$ we can change the norm of the term.
\end{example}

Our goal is similar to one stated above, i.e., to reduce the check of 
rigidity w.r.t. an order to some syntactic condition. Similarly, for the
most general case the emptiness of 
$\mbox{\rm VREL}_>(s)$ is not related to rigidity. The first example
demonstrates that $\mbox{\rm VREL}_>(s) =\emptyset$ is not necessary
for the rigidity, while the second one - that it is not sufficient for it.
\begin{example}
Let $f/1$ be the only functor and $a$ the only constant of the language. 
Let $>$ be defined as follows: $f(X,X) = f(t,t) < f(a,X) = f(X,a)$,
for any term $t$ constructed from the functors and constants of the
underlying language and a set of fresh variables.
In this case $f(X,X)$ is rigid, but $\mbox{\rm VREL}_>(f(X,X))\not =\emptyset$.

Observe that $>$ in this example is not monotone (otherwise, 
$f(a,X) > f(a,a)$ implies $f(f(a,X), f(a,X)) > f(f(a,a), f(a,a))$, that
contradicts the definition above. Additionally, $>$ does not have a
subterm property either, since $f(a,X) > f(f(a,X), f(a,X))$.
\end{example}

\begin{example}
\label{example:empty:non:rigid}
Let $>$ be defined as $f(X,Y) = f(a,Y) = f(f(\ldots),Y) =
f(X,a) = f(X,f(\ldots)) = f(X,X)$.
Then $\mbox{\rm VREL}_>(f(X,Y)) =\emptyset$.
Observe, that this does not imply, for example,
that $\mbox{\rm VREL}_>(f(a,Y)) = \emptyset$. This allows us to define
$f(X,Y) < f(a,a)$, thus, making $f(X,Y)$ non-rigid.

Similarly to the example above, $>$ is not monotone and does not
have a subterm properties. Indeed, if $>$ was monotone
$f(X, f(X,Y)) < f(X,f(a,a))$ was obtained, while both of the terms can be 
obtained by one step replacement from $f(X,Y)$, thus providing a contradiction
to monotonicity. In the same way $f(X,f(a,a)) < f(a,a)$ contradicts the
subterm property. 
\end{example}

We start by studying the rigidity under monotonicity and subterm property
assumptions of $>$. Later on these assumptions will be relaxed.  

\begin{lemma}
\label{rigid:empty:1}
Let $>$ be a monotone order, having a subterm property, on the set of terms 
$S$. Then, if $>$ is rigid on $s\in S$ then $\mbox{\rm VREL}_>(s) = \emptyset$.
\end{lemma}
\begin{proof}
Assume that $\mbox{\rm VREL}_>(s) \not= \emptyset$. This means that there
exists an occurrence $X_{(i)}$ of a variable $X$ and a replacement 
$\{t\rightarrow X_{(i)}\}$, such that $s\{t\rightarrow X_{(i)}\} \not =s$.
On the other hand, $>$ is rigid on $s$. Thus, the replacement cannot be
extended to a substitution $\theta$, such that 
$s\theta = s\{t\rightarrow X_{(i)}\}$. This means, that $s$ is non-linear
in it variables, i.e., $X$ appears among the variables of $s$ at least twice.
Let $X_{(n_1)},\ldots, X_{(n_m)}$ be all occurrences of $X$ in $s$. 
$X_{(i)}$ is one of them.

We distinguish the following cases:
\begin{enumerate}
\item $s < s\{t\rightarrow X_{(i)}\}$. Let $s'$ be a term, obtained from $s$
by a simultaneous replacement of $X_{(n_1)},\ldots,X_{(n_m)}$ in $s$ by
$s$. Let $t'$ be a term, obtained from $s$
by a simultaneous replacement of $X_{(n_1)},\ldots,X_{(n_m)}$ in $s$ by
$s\{t\rightarrow X_{(i)}\}$. Then, by the monotonicity of $>$ holds $s' < t'$.

However, since $X$ does not appear in $s$, except for 
$X_{(n_1)},\ldots, X_{(n_m)}$ and all of those, and only them, have been
replaced by a new terms the compositions of the replacements above
are substitutions. This means that there are two substitutions
$\theta_1$ and $\theta_2$, such that $s\theta_1 < s\theta_2$, contradicting
the rigidity of $s$.
\item $s\{t\rightarrow X_{(i)}\} < s$. Similarly to the previous case.
\item $s\{t\rightarrow X_{(i)}\} \| s$. Let $t'$ be a term, obtained from $s$
by a simultaneous replacement of $X_{(n_1)},\ldots,X_{(n_m)}$ in $s$ by
$s\{t\rightarrow X_{(i)}\}$. Then, by the subterm property of $>$
holds that $s\{t\rightarrow X_{(i)}\} < t'$.
By the same reasoning as above $t'$ is an instance of $s$. Thus,
$s$ is equal to it w.r.t. $>$ (rigidity), i.e., 
$s\{t\rightarrow X_{(i)}\} < s$ providing a contradiction to the 
incomparability.
\end{enumerate}
\end{proof}

\begin{lemma}
\label{empty:rigid:1}
Let $>$ be an order, having a subterm property, on the set of terms 
$S$. Then, if for some $s\in S$ holds that $\mbox{\rm VREL}_>(s) = \emptyset$,
$s$ is ground.
\end{lemma}
\begin{proof}
Assume for the sake of contradiction, that $s$ is not ground. 
Thus, $s$ has at least one variable occurrence say $X_{(i)}$. 
Then, $s\{s\rightarrow X_{(i)}\} > s$, contradicting that
$\mbox{\rm VREL}_>(s) = \emptyset$.
\end{proof}

As we have already pointed out above every ground term is rigid, allowing to
conclude.
\begin{corollary}
Let $>$ be a monotone order, having a subterm property on the set of terms
$s$. Then, the following statements are equivalent:
\begin{enumerate}
\item $>$ is rigid on some $s\in S$
\item $\mbox{\rm VREL}_>(s) = \emptyset$
\item $s$ is ground
\end{enumerate}
\end{corollary}

Thus, if $>$ has monotonicity and subterm properties the notion of rigidity
deteriorates to groundness. As we are going to see this is not necessary the
case if the restrictions on $>$ are relaxed.

In more general setting, presented in Section~\ref{section:weaker:notions}
some correspondence between argument positions possessing monotonicity and 
subterm properties and $\mbox{\rm VREL}_>$ should be established. 

\begin{definition}
Let $s$ be a term, and let $>$ be a partially monotone order, having
a partial subterm property. Let $\mbox{\rm VarInst}(s)$ be the set of all
variable instances of $s$. Then, we denote by $\mbox{\rm M}_>(s)$ and
by $\mbox{\rm S}_>(s)$ the following sets:
\[\mbox{\rm M}_>(s) = \{X_{(j)}| X_{(j)}\in \mbox{\rm VarInst}(s),\;\forall i\;\;\chi^{\cal M}_s(X_{(i)}) = 1\}\]
\[\mbox{\rm S}_>(s) = \{X_{(j)}| X_{(j)}\in \mbox{\rm VarInst}(s),\;\forall i\;\;\chi^{\cal S}_s(X_{(i)}) = 1\}\]
\end{definition}

\begin{lemma}
\label{lemma:subset}
\begin{enumerate}
\item $\mbox{\rm S}_>(s)\subseteq \mbox{\rm VREL}_>(s)$
\item If there exist $t_1$ and $t_2$, such that $t_1 > t_2$, then 
$\mbox{\rm M}_>(s)\cup \mbox{\rm S}_>(s)\subseteq \mbox{\rm VREL}_>(s)$
\end{enumerate}
\end{lemma}
\begin{proof}
\begin{enumerate}
\item Let $X_{(i)}\not\in \mbox{\rm VREL}_>(s)$. This means that for any term
$t$, $s\{t\rightarrow X_{(i)}\} = s$. In particular, 
$s\{s\rightarrow X_{(i)}\} = s$. Clearly, $s$ is a subterm of 
$s\{s\rightarrow X_{(i)}\}$. Thus, by Lemma~\ref{sub:lemma} 
$\chi^{\cal S}_{s\{s\rightarrow X_{(i)}\}}(s) = 0$, i.e., 
$\chi^{\cal S}_s(X_{(i)}) = 0$. 
Thus, $\mbox{\rm S}_>(s)\subseteq \mbox{\rm VREL}_>(s)$.
\item
Let $t_1 > t_2$. If $X_(i)\in \mbox{\rm M}_>(s)$, then
$s\{t_1\rightarrow X_{(i)}\} > s\{t_2\rightarrow X_{(i)}\}$ holds.
Thus, at least one of those terms is not equal to $s$ and 
$X_{(i)}\in \mbox{\rm VREL}_>(s)$. Hence, 
$\mbox{\rm M}_>(s)\subseteq \mbox{\rm VREL}_>(s)$, proving the second statement
of the lemma as well.
\end{enumerate}
\end{proof}

Summarising the definitions introduced so far, an order $>$ maps each
functor $f/n$ (predicate $p/n$) to a pair of sets ${\cal M}_{f/n}$
and ${\cal S}_{f/n}$ (${\cal M}_{p/n}$ and ${\cal S}_{p/n}$) that are defined 
by $>$. 

Now we are able to reformulate the results, stated previously only for orders,
that have a monotonicity and subterm properties 
state them for orders, that are partially monotone and have a partial subterm 
property.

\begin{lemma}
\label{rigid:empty}
Let $>$ be a partially monotone order, having a partial subterm property, 
on the set of terms $S$. Then, if $>$ is rigid on $s\in S$ 
then $\mbox{\rm M}_>(s)\cap\mbox{\rm S}_>(s) = \emptyset$.
\end{lemma}
\begin{proof}(Sketch)
Let $\mbox{\rm M}_>(s)\cap\mbox{\rm S}_>(s)$ be non-empty and let
$X_{(i)}\in \mbox{\rm M}_>(s)\cap\mbox{\rm S}_>(s)$.
Then, Lemma~\ref{mon:lemma} and Lemma~\ref{sub:lemma} allow to conclude
monotonicity and subterm properties for the argument positions occupied by
the instances of $X$ and to mimic the proof of Lemma~\ref{rigid:empty:1}.
\end{proof}

If comparing this lemma with Lemma~\ref{rigid:empty:1} we see that
the assumptions of Lemma~\ref{rigid:empty:1} are {\em more restrictive}
and the conclusion is {\em stronger} than of the recent lemma.

Unfortunately, the second direction of the claim, analogous to 
Lemma~\ref{empty:rigid:1} does not hold. 
Example~\ref{example:empty:non:rigid} illustrates this (under assumption 
that ${\cal M}_{f/2} = {\cal S}_{f/2} = \emptyset$).

\begin{definition}
Let $>$ be a partially monotone order, having a partial subterm property.
Then, the term $s$ is called {\em pseudo-rigid w.r.t. $>$} if 
\begin{itemize}
\item for every substitution $\theta$, 
\item and for every $X\in \mbox{\sl Dom}(\theta)$, such that
\item for every occurrence $X_{(i)}$ of $X$ holds $\chi^{\cal M}_s(X_{(i)}) = 0$
and $\chi^{\cal S}_s(X_{(i)}) = 0$
\[s\theta = s\]
\end{itemize}
\end{definition}

\begin{example}
\label{example:pseudo-rigidity}
Term $[X,Y]$ is pseudo-rigid w.r.t. the list-size based norm.
\end{example}

\begin{lemma}
\label{pseudo-rigidity:rigidity}
Let $>$ be a partially monotone order, having a partial subterm property, 
such that there exist $t_1$ and $t_2$ for which $t_1 > t_2$ holds. 
If $\mbox{\rm VREL}_>(s) = \emptyset$ and $s$ is pseudo-rigid w.r.t. $>$,
then $s$ is rigid w.r.t. $>$.
\end{lemma}
\begin{proof}
Let $\theta$ be a substitution. If $\theta = \varepsilon$, 
$s\theta = s$ and the proof is done. Otherwise, exists some
$X$, such that $X\in \mbox{\sl Dom}(\theta)$.
\eat{
If there exists an occurrence $X_{(i)}$ of $X$, such that
$\chi^{\cal S}_s(X_{(i)}) = 1$. Then,
$s\{s\rightarrow X_{(i)}\} > s$ holds, since $\chi^{\cal S}_s(X_{(i)}) = 1$.
Thus, contradicting $\mbox{\rm VREL}_>(s) = \emptyset$.

If there exists an occurrence $X_{(i)}$ of $X$, such that
$\chi^{\cal M}_s(X_{(i)}) = 1$. Then, if there are two terms $t_1$
and $t_2$ such that $t_1 > t_2$,
$s\{t_1\rightarrow X_{(i)}\} > s\{t_2\rightarrow X_{(i)}\}$ holds, 
since $\chi^{\cal M}_s(X_{(i)}) = 1$. 
Thus, contradicting $\mbox{\rm VREL}_>(s) = \emptyset$.
}
By Lemma~\ref{lemma:subset} 
$\mbox{\rm M}_>(s)\cup \mbox{\rm S}_>(s) =\emptyset$. Thus,
$\mbox{\rm M}_>(s) = \mbox{\rm S}_>(s) =\emptyset$. This means,
that for any $X\in \mbox{\sl Dom}(\theta)$, 
for all occurrences $X_{(i)}$ holds that
$\chi^{\cal S}_s(X_{(i)}) = \chi^{\cal M}_s(X_{(i)}) = 0$. But now,
the pseudo-rigidity condition is applicable, and $s\theta = s$.
\end{proof}
 
For all the examples to be considered further we assume $>$ to be pseudo-rigid
on $\mbox{\sl Call}(P,S)$ as well as the existence of $t_1$ and $t_2$, such
that $t_1 > t_2$. Under these assumptions Lemma~\ref{pseudo-rigidity:rigidity}
allows us to reduce verifying rigidity to verifying the emptiness of
$\mbox{\rm VREL}_>$. That is, we impose that the ordering is invariant on 
predicate argument positions and functor argument positions that may occur 
with a free variable in ${\mbox {\sl Call}}(P,S)$. 
 
\section{Sufficient condition for termination}
Recall that our goal is to extend the 
sufficient condition of~\cite{DeSchreye:Verschaetse:Bruynooghe}, that impose
the additional requirement of rigidity of the level mapping on the call set,
to the case of general term orders. Except for the notion of rigidity,
we also need interargument relations based on general term orders.

\begin{definition}
Let $P$ be a definite program, $p/n$ a predicate in $P$ and $>$ an order on
$U^E_P$. An {\em interargument relation} is a relation 
$R_p = \{(t_1,\ldots,t_n)\mid \varphi_p(t_1, \ldots, t_n)\}$, where:
\begin{itemize}
\item $\varphi_p(t_1,\ldots,t_n)$ is a formula in a disjunctive normal form
\item each conjunct in $\varphi_p$ is either $s_1 > s_2$, $s_1 =_> s_2$ or $s_1\| s_2$, where
$s_i$ are constructed from $t_1,\ldots,t_n$ by applying functors of $P$.
\end{itemize}
$R_p$ is a {\em valid interargument relation for $p/n$ w.r.t. an order $>$}
if and only if for every $p(t_1,\ldots,t_n)\in \mbox{\sl Atom}_P\;\mbox{\rm : if}\;\;
P\models p(t_1,\ldots,t_n)$ then $(t_1,\ldots,t_n)\in R_p$.
\end{definition}

\begin{example}
Consider the following program.
\begin{eqnarray*}
&& p(0,[])\\
&& p(f(X),[X|T])\leftarrow p(X,T)
\end{eqnarray*}
The following interargument relations can be considered for $p$:
$\{(t_1, t_2)\mid t_2 > t_1 \vee t_1 =_> t_2\}$, valid w.r.t. an 
ordering imposed by a list-length norm. Recall, that for lists
$\|[t_1|t_2]\|_l = 1 + \|t_2\|_l$, while the list-length 
of other terms is considered to be 0. 
On the other hand, $\{(t_1, t_2)\mid t_1 > t_2 \vee t_1 =_> t_2\}$, valid 
w.r.t. an ordering imposed by a term-size norm.
$\hfill\Box$\end{example}

Using the notion of rigidity we present a sufficient condition for 
term-acceptability w.r.t. a set.

\begin{theorem}(rigid term-acceptability w.r.t. $S$)
\label{rigid:acceptability}
Let $S$ be a set of atomic queries and $P$ be a definite program.
Let $>$ be an order on $U^E_P$ and for each predicate $p$ in $P$, let $R_p$ be
a valid interargument relation for $p$ w.r.t. $>$. If there exists
a well-founded extension $\succ$ of $>$ to $U^E_P \cup B^E_P$, which is rigid on 
$\mbox{\sl Call}(P,S)$ such that
\begin{itemize}
\item for any clause $H\leftarrow B_1,\ldots,B_n \in P$, and
\item for any atom $B_i$ in its body, such that 
$\mbox{\sl rel}(B_i)\simeq \mbox{\sl rel}(H)$,
\item for substitution $\theta$, such that the arguments of the atoms in
$(B_1,\ldots,B_{i-1})\theta$ all satisfy their associated relations
$R_{\mbox{\sl rel}(B_1)},\ldots, R_{\mbox{\sl rel}(B_{i-1})}$
\end{itemize}
\[H\theta \succ B_i\theta\]

then $P$ is term-acceptable w.r.t. $S$
\end{theorem}
\begin{proof}
Suppose the above condition is satisfied for $P$. Take any 
$A\in \mbox{\sl Call}(P,S)$ and any clause $A'\leftarrow B_1,\ldots,B_n$
such that $\mbox{\rm mgu}(A,A') = \theta$ exists. Suppose, that
$B_i$ is a body atom, such that $\mbox{\sl rel}(B_i)\simeq \mbox{\sl rel}(A)$
and that $\sigma$ is a computed answer substitution for 
$\leftarrow (B_1, \ldots, B_{i-1})\theta$. 

Then $A\theta$ is identical to $A'\theta$, and thus, 
$A\theta\sigma$ is identical to $A'\theta\sigma$. Since $\succ$ is
rigid on $\mbox{\sl Call}(P,S)$ and $A\in \mbox{\sl Call}(P,S)$,
$A'\theta\sigma =_\succ A$.

Finally, since $\sigma$ is a computed answer substitution 
$P\models B_j\theta\sigma$, for all $j < i$. Thus, by definition of valid
interargument relation the arguments of $B_j\theta\sigma$ satisfy 
$R_{\mbox{\sl rel}(B_j)}$,
for all $j < i$. Thus, by the rigid term-acceptability assumption 
$A'\theta\sigma \succ B_i\theta\sigma$. Combined with 
$A'\theta\sigma =_\succ A$, we get $A > B_i\theta\sigma$.
\end{proof} 

We continue the analysis of Example~\ref{example:permute}.
\begin{example}
Let $>$ be a well-founded ordering on $B^E_P \cup U^E_P$, such that:
\begin{itemize}
\item for all terms $t_1, t_{21}$ and $t_{22}$:
${\mbox{\sl permute}}(t_1, t_{21}) =_> {\mbox{\sl permute}}(t_1, t_{22})$.
\item for all terms $t_{11}, t_{12}, t_2, t_{31}, t_{32}$:
${\mbox{\sl delete}}(t_{11}, t_{2}, t_{31}) =_> {\mbox{\sl delete}}(t_{12}, t_{2}, t_{32})$.
\item for all terms $t_{11}, t_{12}$ and $t_2$:
$[t_{11}| t_2] =_> [t_{12}| t_2]$.
\end{itemize}

That is, we impose that the ordering is invariant on predicate argument 
positions and functor argument positions that may occur with a free variable in
${\mbox {\sl Call}}(P,S)$. Furthermore, we impose that $>$ has the subterm
and monotonicity properties at all remaining predicate or functor argument
positions. 

First we investigate the rigidity of $>$ on ${\mbox {\sl Call}}(P,S)$, namely:
$G\theta =_> G$ for any $G\in{\mbox {\sl Call}}(P,S)$ and any $\theta$. Now
any effect that the application of $\theta$ to $G$ may have on $G$ needs to be 
through the occurrence of some variable in $G$. However, because we imposed that
$>$ is invariant on all predicate and functor argument positions that may 
possibly contain a variable in some call, $G\theta =_> G$.

Associate with ${\mbox{\sl delete}}$ the interargument relation 
$R_{\mbox{\sl delete}} = \{(t_1, t_2, t_3)\mid t_2 > t_3\}$.
First, we verify that this interargument relationship is valid. Note, that
an interargument relationship is valid whenever it is a model for its 
predicate.
Thus, to check whether $R_{\mbox{\sl delete}}$ is valid, 
$T_P(R_{\mbox{\sl delete}})\subseteq R_{\mbox{\sl delete}}$ is checked.
For the non-recursive clause of ${\mbox{\sl delete}}$ the inclusion follows
from the subset property of $>$, while for the recursive one, from the 
monotonicity of it.

Then, consider the recursive clauses of the program. 
\begin{itemize}
\item ${\mbox{\sl permute}}$. If ${\mbox{\sl delete}}(El,L,L1)\theta$ 
satisfies $R_{\mbox{\sl delete}}$, then $L\theta > L1\theta$.
By the monotonicity, ${\mbox{\sl permute}}(L,T)\theta > {\mbox{\sl permute}}(L1,T)\theta$. By the property stated above,
 ${\mbox{\sl permute}}(L,[El|T])\theta =_> {\mbox{\sl permute}}(L,T)\theta$.
Thus, the desired decrease 
${\mbox{\sl permute}}(L,[El|T])\theta > {\mbox{\sl permute}}(L1,T)\theta$
holds.
\item ${\mbox{\sl delete}}$. 
By the properties of $>$ stated above:
${\mbox{\sl delete}}(X,[H|T],[H|T1]) > {\mbox{\sl delete}}(X,T,[H|T1])$
and ${\mbox{\sl delete}}(X,T,[H|T1]) =_> {\mbox{\sl delete}}(X,T,T1)$.
Thus,
${\mbox{\sl delete}}(X,[H|T],[H|T1]) > {\mbox{\sl delete}}(X,T,T1)$. 
\end{itemize}

We have shown that all the conditions of Theorem~\ref{rigid:acceptability}
are satisfied, and thus, $P$ is term-acceptable w.r.t. $S$. By Theorem~\ref{taset:term}, $P$ terminates for all queries in $S$.

Observe, that we do not need to construct the actual order, but only
to prove that there is some, that meets all the requirements posed. In this 
specific case, the requirement of subterm and monotonicity on the remaining
argument positions is satisfiable.
$\hfill\Box$\end{example}

\section{The results for standard acceptability}
In this section we briefly discuss some of the results we obtained in 
generalising the acceptability notion of~\cite{Apt:Pedreschi:Studies}.
Since these results are weaker than those presented in the previous
section, we do not elaborate on them in full detail. In particular, we do not
recall the definitions of well-moded programs and goals, nor those of
simply moded programs and goals, that we use below, 
but instead refer to~\cite{Apt:Book}, respectively~\cite{Apt:Etalle}.
Below, we assume that in-output modes for the program and goal are given.
For any atom $A$ and a mode $m_A$ for $A$, we denote by $A^{\mbox{\sl inp}}$ 
the atom obtained from $A$ by removing all output arguments. 
E.g., let $A = p(f(2), 3, X)$ and $m_A = p(\mbox{\sl in},\mbox{\sl in}, \mbox{\sl out})$, then $A^{\mbox{\sl inp}} = p(f(2), 3)$.

\begin{definition}
Let $>$ be an order relation on $B^E_P$. We say that $>$ is {\em output-independent\/} if for any two moded atoms $A$ and $B$: $A^{\mbox{\sl inp}} = B^{\mbox{\sl inp}}$ implies $A =_> B$.
\end{definition}

For well-moded programs, term-acceptability in the style of~\cite{Apt:Pedreschi:Studies} can now be defined as follows.

\begin{definition}
Let $P$ be a well-moded program, $>$ an output-independent well-founded order 
and $I$ a model for $P$. The program $P$ is called {\em term-acceptable 
w.r.t.\ $>$ and $I$\/} if for all $A\leftarrow B_1,\ldots,B_n$ in $P$ and all 
substitutions $\theta$, such that $(A\theta)^{\mbox{\sl inp}}$ and
$B_1\theta,\ldots,B_{i-1}\theta$ are ground and 
$I\models B_1\theta\wedge\ldots B_{i-1}\theta$ holds: $A\theta > B_i\theta$.
\end{definition}

$P$ is called {\em term-acceptable\/} if it is term-acceptable w.r.t.\ some 
output-independent well-founded order and some model. The following theorem 
states that 
term-acceptability of a well-moded program is sufficient for termination
of well-moded goals w.r.t.\ this program.

\begin{theorem}
\label{term-acc:term}
Let $P$ be a well-moded program, that is term-acceptable w.r.t. an 
output-independent well-founded order $>$ and a model $I$. 
Let $G$ be a well-moded goal, then $G$ LD-terminates.
\end{theorem}

\begin{proof}
We base our proof on the notion of directed sequence.
Let $G$ be non-terminating, i.e., $P\cup \{G\}$ has an infinite derivation.
By Lemma~\ref{Verschaetse:lemma} it has an infinite directed 
derivation as well. Let $G_0, G_1,\ldots$ be this infinite directed derivation.
We denote $G_i = \leftarrow A^i_1,\ldots A^i_{n_i}$ and 
$G_{i+1} = \leftarrow A^{i+1}_1,\ldots A^{i+1}_{n_{i+1}}$. There is
a clause $H\leftarrow B_1,\ldots,B_n$, and substitutions $\sigma$ and 
$\theta$ such that $A^i_1 = H\sigma$, for some $1\leq j\leq n_{i+1}$, 
$A^{i+1}_1 = B_j\sigma\theta$,
$I\models B_1\sigma\theta\wedge\ldots,B_{j-1}\sigma\theta$ 
and $B_1\sigma\theta\wedge\ldots,B_{j-1}\sigma\theta$ is ground. 
Note, that  $(H\sigma)^{\mbox {\sl inp}}$ is ground due to the well-modedness.
The term-acceptability condition implies that
$H\sigma\theta > B_j\sigma\theta$, that is $A^i_1\theta > A^{i+1}_1$. 
Since $P$ and $G$ are well-founded $(A^i_1)^{\mbox {\sl inp}}$ is ground.
Thus, $(A^i_1)^{\mbox {\sl inp}} = (A^i_1\theta)^{\mbox {\sl inp}}$ and,
by the output-independence of $>$, $A^i_1 =_> A^i_1\theta$.
By transitivity, $A^i_1 > A^{i+1}_1$. Thus, selected atoms of the 
goals in the infinite directed derivation form an infinite decreasing chain
w.r.t.\ $>$, contradicting the well-foundedness of the order.
\end{proof}

Note that if the requirement of well-modedness is relaxed the theorem
does not hold anymore.
\begin{example}
\label{pa:example}
\eat{
\begin{eqnarray*}
&& p(a) \leftarrow q(X)\\
&& q(f(X)) \leftarrow q(X)
\end{eqnarray*}
}
\[p(a) \leftarrow q(X)\;\;\;\;\;q(f(X)) \leftarrow q(X)\]
We assume the modes $p(\mbox{\sl in})$ and $q(\mbox{\sl in})$ to be given.
This program is not well-moded w.r.t.\ the given modes, but it satisfies the
remaining conditions of term-acceptability with respect to the 
following order $>$ on terms
\[p(a) > \ldots > q(f(f(f(a)))) > q(f(f(a))) > q(f(a)) > q(a)\]
and a model $I = \{p(a), q(a), q(f(a)), q(f(f(a))), \ldots\}$.
However, note that the well-founded goal $p(a)$ is non-terminating.
$\hfill\Box$\end{example}

Unfortunately, well-modedness is not sufficient to make the converse to hold.
That is, there is a well-moded program $P$ and a well-moded goal $G$, such that
$G$ is LD-terminating w.r.t.\ $P$, but $P$ is not term-acceptable.  
\begin{example}
Consider the following program
\begin{eqnarray*}
&& p(f(X)) \leftarrow p(g(X))
\end{eqnarray*}
with the mode $p(\mbox{\sl out})$. This program is well-moded, 
the well-moded goal $p(X)$ 
terminates w.r.t.\ this program, but it is not term-acceptable, 
since the required decrease $p(f(X)) > p(g(X))$ violates 
output-independence of $>$.
$\hfill\Box$\end{example}

Intuitively, the problem in the example occured, since some information
has been passed via the output positions, i.e, $P$ is not simply
moded. Indeed, if $P$ is simply-moded,\cite{Apt:Etalle}, the second direction 
of the theorem holds as well.

We start the presentation by a number of useful lemmas.

\begin{lemma} 
\label{suffix:lemma}
Let $P_0,\ldots,P_m$ be directed sequence, such that
$P_i =\leftarrow A^i_1,\ldots,A^i_{n_i}$. Then, for any suffix ${\bf S}$,
exist a sequence of substitutions $\theta_1,\ldots,\theta_m$ and a sequence
of suffices ${\bf R_1},\ldots,{\bf R_m}$ such that
\[\leftarrow A^0_1, {\bf S}; \;\; \leftarrow A^1_1, {\bf R_1}\theta_1, {\bf S}\theta_1; \;\;\ldots; \;\; \leftarrow A^m_{n_m}, {\bf R_m}\theta_1\ldots\theta_m, {\bf S}\theta_1\ldots\theta_m\] is directed.
\end{lemma}

\begin{lemma}
\label{concatenation:lemma}
Let $P_0,\ldots,P_m$ and $Q_0,\ldots,Q_k$ be two directed sequences, such that
$P_m =\leftarrow A_1,\ldots,A_s$ and $Q_0 =\leftarrow B_1,\ldots,B_t$
and $A_1 = B_1$. Then, exists a directed sequence $R_0,\ldots,R_{m+k}$, such 
that the selected atom of $R_0$ is the selected atom of $P_0$, and the selected
atom of $R_{m+k}$ is the selected atom of $Q_k$.
\end{lemma}
\begin{proof} 
We define $R_0,\ldots,R_{m+k}$ as following:
\[R_i = \left\{
\begin{array}{ll}
P_i & \mbox{\rm if $0\leq i\leq m$} \\
T_{i-m} & \mbox{\rm if $m\leq i\leq m+k$}
\end{array}
\right.
\]
where $T_j$ is the $j$-th element in the sequence, generated by
Lemma~\ref{suffix:lemma} for the directed sequence $Q_0,\ldots,Q_k$ with 
${\bf S} = A_2,\ldots,A_s$.

The sequence $R_0,\ldots,R_{m+k}$ is well-defined: if $i = m$, on one hand we 
get that $R_m = P_m$, and on the other hand, $R_m = B_1, {\bf S}$, that is 
$P_m$. For $i\not = m$ only one of those definitions is applicable. 

The requirement of the lemma are clearly fulfilled.
\end{proof}

\begin{corollary}
\label{concatenation:corollary}
Let $P$ be a simply moded program and $Q, Q'$ - simply moded goals.
Let $P_0,\ldots,P_m$ be a directed sequence obtained from one of the 
derivations for $P\cup \{Q\}$ and
and $Q_0,\ldots,Q_k$ be a directed sequence obtained from one of the 
derivations for $P\cup \{Q'\}$. Let also  
$P_m$ be $\leftarrow A_1,\ldots,A_s$, $Q_0$ be $\leftarrow B_1,\ldots,B_t$
and $A_1^{\mbox{\sl inp}} = B_1^{\mbox{\sl inp}}$. 
Then, exists a directed sequence $R_0,\ldots,R_{m+k}$, such 
that the selected atom of $R_0$ is the selected atom of $P_0$, and the selected
atom of $R_{m+k}$ is the selected atom of $Q_k$.
\end{corollary}
\begin{proof}
Since $P_m$ and $Q_0$ are queries in some of the LD-derivations of the simply 
moded queries and of the simply moded program, they are simply moded~\cite{Apt:Etalle}. Thus, the output positions, both of $A_1$ and of $B_1$, are occupied 
by distinct variables. Since $A_1^{\mbox {\sl inp}} = B_1^{\mbox {\sl inp}}$  
we can claim that $A_1 = B_1$, up to variables renaming. Thus, 
Lemma~\ref{concatenation:lemma} becomes applicable (note that we never required
in the lemma that both directed sequences shall originate from the derivations of the same LD-tree), and we can obtain a new
directed sequence as required.
\end{proof}

\begin{theorem}
\label{term:term-acc}
Let $P$ be a well-moded and simply moded program, LD-terminating for any
well-moded and simply-moded goal. Then there exists a
model $I$ and an output-independent well-founded order $>$, 
such that $P$ is term-acceptable w.r.t.\ $I$ and $>$.
\end{theorem}

\begin{proof}
We base the choice of $>$ on the LD-trees. More precisely, we define 
$A>B$ if there is a well-moded and simply moded goal $G$ and there is a
directed sequence $P_0,\ldots,P_m$ in the LD-tree for $P\cup \{G\}$, such that 
the selected atom of $P_0$ is $A_1$ and the selected atom 
of $P_m$ is $B_1$ and $A^{\mbox{\sl inp}} = A_1^{\mbox{\sl inp}}$,
$B^{\mbox{\sl inp}} = B_1^{\mbox{\sl inp}}$.
Let $I$ be a lest Herbrand model of $P$.
We have to prove that:
\begin{enumerate}
\item $>$ is an order relationship, i.e., is irreflexive, asymmetric and transitive;
\item $>$ is output-independent;
\item $>$ is well-founded;
\item $P$ is term-acceptable w.r.t.\ $>$ and $I$
\end{enumerate}
\begin{enumerate}
\item $>$ is an order relationship, that is $>$ is irreflexive, asymmetric
and transitive. 
\begin{enumerate}
\item Irreflexivity. If $A>A$ holds, then exists a directed sequence
$P_0,\ldots,P_k$, such that the selected atom of $P_0$ is $A_1$, the
selected atom of $P_k$ is $A_2$, $A_1^{\mbox{\sl inp}} = A^{\mbox{\sl inp}}$ 
and $A_2^{\mbox{\sl inp}} = A^{\mbox{\sl inp}}$. 
By repetitive application of Corollary~\ref{concatenation:corollary} 
an infinite branch is build and the contradiction to the finiteness
of the LD-tree is obtained.
\item Asymmetry. If $A>B$ holds, then exists a directed sequence
$P_0,\ldots,P_k$, such that the selected atom of $P_0$ is $A_1$, the
selected atom of $P_k$ is $B_1$, $A^{\mbox{\sl inp}} = A_1^{\mbox{\sl inp}}$,
$B^{\mbox{\sl inp}} = B_1^{\mbox{\sl inp}}$.
If $B>A$ holds, then exists a directed sequence
$Q_0,\ldots,Q_m$, such that the selected atom of $Q_0$ is $B_2$, the
selected atom of $Q_m$ is $A_2$, $A^{\mbox{\sl inp}} = A_2^{\mbox{\sl inp}}$,
$B^{\mbox{\sl inp}} = B_2^{\mbox{\sl inp}}$.
By repetitive application of Corollary~\ref{concatenation:corollary} 
an infinite branch is build and the contradiction to the finiteness
of the LD-tree is obtained.
\item Transitivity. If $A>B$ holds, then exists a directed sequence
$P_0,\ldots,P_k$, such that the selected atom of $P_0$ is $A_1$, the
selected atom of $P_k$ is $B_1$, $A^{\mbox{\sl inp}} = A_1^{\mbox{\sl inp}}$,
$B^{\mbox{\sl inp}} = B_1^{\mbox{\sl inp}}$.
 If $B>C$ holds, then exists a directed sequence
$Q_0,\ldots,Q_m$, such that the selected atom of $Q_0$ is $B_2$, the
selected atom of $Q_m$ is $C_2$,$B^{\mbox{\sl inp}} = B_2^{\mbox{\sl inp}}$,
$C^{\mbox{\sl inp}} = C_2^{\mbox{\sl inp}}$.
By applying the Corollary~\ref{concatenation:corollary} 
new directed sequence is build, such that the selected atom of its first
element is $A_1$ and the selected atom of its last element is $C_2$. By 
definition of $>$ holds that $A>C$.
\end{enumerate}
\item $>$ is output-independent. Assume that there are two atoms $A$ and $B$,
such that $A^{\mbox{\sl inp}} = B^{\mbox{\sl inp}}$, but $A>B$.
Then exists a directed sequence $P_0,\ldots,P_k$, such that the selected 
atom of $P_0$ is $A_1$, the selected atom of $P_k$ is $B_1$, 
$A_1^{\mbox{\sl inp}} = A^{\mbox{\sl inp}}$ and 
$B_1^{\mbox{\sl inp}} = B^{\mbox{\sl inp}}$. However,
$B_1^{\mbox{\sl inp}} = A^{\mbox{\sl inp}}$.
Thus, by repetitive application of Corollary~\ref{concatenation:corollary} 
an infinite branch is build and the contradiction to the finiteness
of the LD-tree is obtained.
\item $>$ is well-founded. Assume that there is an infinitely decreasing
chain $A_1 > A_2 >\ldots$. This means that there is an infinite directed
sequence in the tree (concatenation of infinitely many finite ones), 
contradicting the finiteness of the tree.
\item $P$ is term-acceptable w.r.t.\ $>$ and $I$. Let 
$A\leftarrow B_1,\ldots,B_n$. 
Let $\theta$ be a substitution, such that $(A\theta)^{\mbox {\sl inp}}, 
B_1\theta,\ldots, B_{i-1}\theta$ are 
ground and $I\models B_1\theta\wedge\ldots\wedge B_{i-1}\theta$.
The goal $\leftarrow A\theta$ is a well-moded goal, however, it is not
necessary simply moded. Thus, we define a new goal $A'$ such that it
will coincide with $A\theta$ on its input positions, and its output positions
will be occupied by a linear set of variables.

More formally, let $\theta_1$ be $\theta$ restricted to 
{\sl Var}$(A)^{\mbox{\sl inp}}$. Then 
$(A\theta_1)^{\mbox{\sl inp}} = (A\theta)^{\mbox{\sl inp}}$, and thus,
$(A\theta_1)^{\mbox{\sl inp}}$ is ground, while $(A\theta_1)^{\mbox{\sl out}} 
= A^{\mbox{\sl out}}$, and thus, $(A\theta_1)^{\mbox{\sl out}}$ is a linear
sequence of variables. Summing up, $\leftarrow A\theta_1$ is well-moded
and simply moded goal. Thus, it terminates w.r.t.\ $P$ and its derivations
has been considered while defining $>$.

By definition of $\theta_1$ exists some substitution $\sigma$, such that
$\theta = \theta_1\sigma$. Thus, $B_1\theta,\ldots, B_{i-1}\theta =
(B_1\theta_1,\ldots,B_{i-1}\theta_1)\sigma$. Since $I$ is a 
least Herbrand model $\sigma$ is a correct answer substitution of 
$B_1\theta_1,\ldots,B_{i-1}\theta_1$ and, since $P$ and 
$\leftarrow B_1\theta_1,\ldots,B_{i-1}\theta_1$
are well-moded (the later as the LD-resolvent of the well-moded clause
and the well-moded goal), $\sigma$ is a computed answer substitution as 
well~\cite{Apt:Book}. Thus, the next goal to be considered in the derivation
is $\leftarrow B_i\theta_1\sigma,\ldots,B_n\theta_1\sigma$. This is a directed
descendant of $A\theta_1$, thus, by definition of $>$, 
$A\theta_1 > B_i\theta_1\sigma$. By definition of $\sigma$, 
$B_i\theta_1\sigma = B_i\theta$, and by the definition of $\theta_1$,
$(A\theta_1)^{\mbox{\sl inp}} = (A\theta)^{\mbox{\sl inp}}$. 
Thus, $A\theta > B_i\theta$.
\end{enumerate}
\end{proof}

We have introduced so far two different notions of term-acceptability:
notion of term-acceptability w.r.t. a set of queries and the notion
of term-acceptability (w.r.t. an order and a model). We study the
relationship between these two notions for well-moded and simply-moded
programs and goals.
 
Observe, that the well-modedness and simply modedness are not sufficient
to reduce the term-acceptability w.r.t. a set to the term-acceptability.
\begin{example}
\begin{eqnarray*}
&& q\\
&& p\leftarrow p
\end{eqnarray*}
This program is not term-acceptable, but it is term-acceptable w.r.t.
a set $\{q\}$.
\end{example}

In order to eliminate this kind of examples we assume also in 
Theorem~\ref{taset:term-acc}, that for every clause in $P$ there is a goal in 
$\mbox{\sl Call}(P,S)$, that can be unified with its head.

Once more we precede the theorem with a small lemma.
\begin{lemma}
\label{wmsm:lemma}
Let $P$ be a well-moded and simply moded program, and $S$ be a set of
well-moded and simply moded goals. Then, given an output-independent $>$, for
any clause goal $A\in\mbox{\sl Call}(P,S)$ and any 
$A'\leftarrow B_1,\ldots,B_n$, s.t. $\mbox{\rm mgu}(A,A') = \theta$ exists, 
holds that $A =_> A'\theta$.
\end{lemma}
\begin{proof}
If $G\in S$ and $P$ are well-moded then $A$ is well-moded as 
well~\cite{Apt:Book}.
Analogously, $A$ is simply moded~\cite{Apt:Etalle}. Thus, the input positions 
of $A$ are ground and the output positions of $A$ are occupied by distinct
variables. Therefore, $\theta$ cannot affect the input positions of $A$, and
$A^{\mbox{\sl inp}} = (A\theta)^{\mbox{\sl inp}}$, i.e., $A =_> A\theta$. 
Since $A\theta$ is identical to $A'\theta$, $A =_> A'\theta$ holds.
\end{proof}

\begin{theorem}
\label{taset:term-acc}.
Let $S$ be a set of well-moded and simply moded goals,
$P$ be a well-moded and simply moded program, such that for every 
clause in $P$ there is a goal in $\mbox{\sl Call}(P,S)$, that can be unified 
with its head. Then, given an output-independent $>$,
if $P$ is term-acceptable w.r.t. $S$ and $>$, then $P$ is term-acceptable
w.r.t. some well-founded order $\succ$ and least Herbrand model of $P$.
\end{theorem}
\begin{proof}
We define $\succ$ in the following way:
\[
t_1\succ t_2\;\;{\rm if}\;\;\left\{
\begin{array}{l}
rel(t_1)\sqsupset rel(t_2)\\
rel(t_1)\simeq rel(t_2)\;\;{\rm and}\;\;t_1 > t_2
\end{array}
\right.
\]
The properties of $\succ$ follow from the corresponding properties of 
$>$ and $\sqsupset$.

Let $A'\leftarrow B_1,\ldots,B_n$ be a clause.
We have to prove that for any substitution $\gamma$, such that 
$(A'\gamma)^{\mbox{\sl inp}}$ and $(B_1, \ldots, B_{i-1})\gamma$ are ground and
$I\models B_1\gamma\wedge \ldots\wedge B_{i-1}\gamma$ holds that
$A'\gamma \succ B_i\gamma$.

Then, by the property of $P$ stated above
there is some $A\in \mbox{\sl Call}(P,S)$, unifiable with
$A'$. Let $\theta$ be the most general unifier of $A$ and
$A'$. By Lemma~\ref{wmsm:lemma} $A =_> A'\theta$. On the other hand, 
$(A'\theta)^{\mbox{\sl inp}} = (A'\gamma)^{\mbox{\sl inp}}$. On the other
hand, the output argument positions of $A'\theta$ are occupied by a sequence
of distinct variables (simply modedness of $P$ and $A$ - we can always assume
that fresh variants of the clauses are considered~\cite{Apt:Etalle}). Thus,
there is some $\sigma$, such that $\gamma = \theta\sigma$.

$I\models B_1\gamma\wedge \ldots\wedge B_{i-1}\gamma$. Since $I$ is the
least Herbrand model, $\gamma$ is a correct answer substitution
for $\leftarrow (B_1, \ldots, B_{i-1})$. Since $P$ and any $G\in S$ 
are well-moded $\gamma$ is also computed answer substitution. Thus, for the 
goal $\leftarrow (B_1, \ldots, B_{i-1})\theta$ the computed answer substitution
will be $\sigma$. 

\begin{itemize}
\item $\mbox{\sl rel}(B_i)\simeq \mbox{\sl rel}(A')$. 
Then, by the definition of term-acceptability w.r.t. a set, 
$A > B_i\theta\sigma$. However, $A =_> A'\gamma$ and $B_i\theta\sigma$
is $B_i\gamma$. Thus, by claiming $A'\gamma > B_i\gamma$, and
$A'\gamma \succ B_i\gamma$.

\item $\mbox{\sl rel}(B_i)\not\simeq \mbox{\sl rel}(A')$. This means that
$\mbox{\sl rel}(A')\sqsupset \mbox{\sl rel}(B_i)$, i.e., 
$A'\gamma\succ B_i\gamma$.
\end{itemize}
\end{proof}

To conclude, we briefly discuss why it is difficult to extend the notions of term-acceptability to the non well-moded case, using a notion of  
boundedness, as it was done for standard acceptability~\cite{Apt:Pedreschi:Studies}. In acceptability based on level mappings, boundedness ensures that the 
level mapping of a (non-ground)
goal can only increase up to some finite bound when the goal becomes more
instantiated. Observe that every ground goal is trivially bounded. 

One particular possible generalisation
of boundedness to term-orderings, which is useful for maintaining most of our 
results, is:

An atom $A$ is {\em bounded} with respect to an ordering $<$, if there exists
an atom $C$ such that for all ground instances $A\theta$ of $A$: $A\theta < C$,
and $\{B\in B^E_P\mid B < C\}$ is finite. 

Such a definition imposes constraints which are very similar to the ones
imposed by standard boundedness in the context of level mappings. However, one
thing we loose is that it is no longer generalisation of groundness. 
Consider an atom $p(a)$ and assume that our language contains a functor $f/1$ 
and a constant $b$. Then one particular well-founded ordering is
\[p(a) > \ldots > p(f(f(b))) > p(f(b)) > p(b).\]
So, $p(a)$ is not bounded with respect to this ordering.

Because of such complications, we felt that the rigidity-based results
of the previous section are the prefered generalisations to general term 
orders.

\section{Towards automation of the approach}
In this section we present an approach leading towards 
automatic verification of the 
term-acceptability condition. The basic idea for the 
approach is inspired on the ``constraint based'' termination analysis proposed
in~\cite{Decorte:DeSchreye:Vandecasteele}. We start off from the conditions
imposed by term-acceptability, and systematically reduce these conditions
to more explicit constraints on the objects of our search: the order $>$ 
and the interargument relations, $R_p$, or model $I$.

The approach presented below 
has been applied successfully to a number of examples that
appear in the literature on termination, such as different versions of
{\sl permute}~\cite{Arts:Zantema:95,Krishna:Rao,Decorte:DeSchreye:Vandecasteele}, {\sl dis-con}~\cite{DeSchreye:Decorte:NeverEndingStory}, {\sl transitive closure}~\cite{Krishna:Rao}, {\sl add-mult}~\cite{Plumer:Book}, {\sl combine}, {\sl reverse}, {\sl odd-even}, {\sl at\_least\_double} and {\sl normalisation}~\cite{Decorte:DeSchreye:Vandecasteele}, {\sl quicksort} program~\cite{Sterling:Shapiro,Apt:Book}, {\sl derivative}~\cite{DM79:cacm}, {\sl distributive law}~\cite{Dershowitz:Hoot}, {\sl boolean ring}~\cite{Hsiang},
{\sl flatten}~\cite{Arts:PhD}.

In the remainder of the paper, we explain the approach using some 
of these examples. 

We start by showing how the analysis of Example~\ref{example:permute}, 
presented before, can be performed systematically. We stress the main steps 
of an algorithm.

\begin{example}
$>$ should be rigid on ${\mbox {\sl Call}}(P,S)$. To enforce the rigidity,
$>$ should ignore all argument positions in atoms in ${\mbox {\sl Call}}(P,S)$
that might be occupied by free variables, i.e., the second argument
position of ${\mbox {\sl permute}}$ and the first and the third argument 
positions of ${\mbox {\sl delete}}$. Moreover, since the first argument
of ${\mbox {\sl permute}}$ and the second argument of ${\mbox {\sl delete}}$
are general nil-terminated lists, the first argument of $./2$ should be 
ignored as well. 

The $>$-decreases imposed in the term-acceptability w.r.t. a set $S$ are:
\begin{eqnarray*}
&& {\mbox {\sl delete}}(X, [H|T], [H|T1])\theta > {\mbox {\sl delete}}(X, T, T1)\theta \\
&& {\mbox {\sl delete}}(El, L, L_1)\theta\;\;{\mbox {\rm satisfies}}\;\;R_{\mbox {\sl delete}}\;\;{\mbox {\rm implies}}\\
&& \hspace{1.0cm}{\mbox {\sl permute}}(L,[El|T])\theta > {\mbox {\sl permute}}(L_1,T)\theta
\end{eqnarray*}

Each of these conditions we simplify by replacing the predicate argument 
positions that should be ignored by some arbitrary term $t$. 
The following conditions are obtained:
\begin{eqnarray}
&& {\mbox {\sl delete}}(t, [H|T]\theta, t) > {\mbox {\sl delete}}(t, T\theta, t) \\
&& {\mbox {\sl delete}}(El, L, L_1)\theta\;\;{\mbox {\rm satisfies}}\;\;R_{\mbox {\sl delete}}\;\;{\mbox {\rm implies}} \nonumber\\
&& \hspace{1.0cm}{\mbox {\sl permute}}(L\theta, t) > {\mbox {\sl permute}}(L_1\theta, t)
\end{eqnarray}

Observe that this only partially deals with the requirements that the 
rigidity conditions expressed above impose: rigidity on functor arguments
(the first argument of $./2$ should be invariant w.r.t. the order) is not 
expressed. We keep track of such constraints implicitly, and only verify them
at a later stage when additional constraints on the order are derived.

For each of the conditions (1) and (2), we have two options on how to enforce it:

Option 1): The decrease required in the condition can be achieved by imposing
some property on $>$, which is consistent with the constraints that were 
already imposed on $>$ before.

In our example, condition (1) is satisfied by imposing the subterm property 
for the second argument of $./2$ and monotonicity on the second argument
of ${\mbox {\sl delete}}$. The second argument of $./2$ does not belong to
a set of functor argument positions that should be ignored.
Then, $[t_1|t_2] > t_2$ holds for any terms $t_1$ and $t_2$, and by the
monotonicity of $>$ in the second argument of ${\mbox {\sl delete}}$ 
(1) holds. 

In general we can select from a bunch of term-order properties, or even 
specific term-orders, that were proposed in the literature. 

Option 2): The required decrease is imposed as a constraint on the 
interargument relation(s) $R$ of the preceding atoms.

In the ${\mbox {\sl permute}}$ example, the decrease 
${\mbox {\sl permute}}(L\theta, t) > {\mbox {\sl permute}}(L_1\theta, t)$ 
cannot directly be achieved by 
imposing some constraint on $>$. Thus, we impose that the underlying decrease
$L\theta > L_1\theta$ should hold for the intermediate body atoms
(${\mbox {\sl delete}}(El, L, L_1)\theta$) that satisfy the interargument
relation $R_{\mbox {\sl delete}}$.

Thus, in the example, the constraint is that 
$R_{\mbox {\sl delete}}$ should be such that for all
${\mbox {\sl delete}}(t_1, t_2, t_3)$ that satisfy $R_{\mbox {\sl delete}}$: 
$t_2 > t_3$. As we have observed, the interargument relation is valid if it
forms a model for its predicate. Thus, one way to constructively verify that 
a valid interargument relation $R_{\mbox {\sl delete}}$ exists,
such that the property $t_2 > t_3$
holds for ${\mbox {\sl delete}}(t_1, t_2, t_3)$ atoms is to simply impose
that $M = \{{\mbox {\sl delete}}(t_1, t_2, t_3)\mid t_2 > t_3\}$ itself is
a model for the ${\mbox {\sl delete}}$ clauses in the program. 

So our new constraint on $R_{\mbox {\sl delete}}$ is that it should 
include $M$. Practically we
can enforce this by imposing that $T_P(M)\subseteq M$ should hold. 
\eat{
in~\cite{Serebrenik:DeSchreye:cw291}, this reduces to the constraints 
``$[t_1|t_2] > t_2$'' and ``$t_2 > t_3$ implies $[t|t_2] > [t|t_3]$''. 
}
That is, the following constraints are obtained:
\begin{eqnarray*}
&& [t_1|t_2] > t_2\\
&& t_2 > t_3\;\;\mbox{\rm implies}\;\;[t|t_2] > [t|t_3]
\end{eqnarray*}

These are again fed into our
Option 1) step, imposing a monotonicity property on the second argument of 
$./2$ for $>$
. At this point the proof is complete.
$\hfill\Box$\end{example}

Recall, that we do not need to construct actually the order, but only to prove
that there is some, that meets all the requirements posed.

The previous example does not illustrate the approach in full generality.
It might happen that more than one intermediate goal preceded the recursive 
atom in the body of the clause. In this case we refer to the whole conjunction as to 
``one'' subgoal. Formally, given a sequence of intermediate body atoms 
$B_1,\ldots,B_n$ a (generalised) clause 
$B_1,\ldots,B_n\leftarrow B_1,\ldots,B_n$ is constructed and one 
step of unfolding is performed on each atom in its body, 
producing a generalised program $P'$. 
\begin{example}
The following is the version of the {\sl permute} program that appeared 
in~\cite{Krishna:Rao}.
\[
\begin{array}{ll}
\mbox{\sl perm}([],[]). & \mbox{\sl ap}_1([],L,L).\\
\mbox{\sl perm}(L,[H|T])\leftarrow & \mbox{\sl ap}_1([H|L1],L2,[H|L3])\leftarrow\\
\hspace{0.5cm} \mbox{\sl ap}_2(V,[H|U],L), &\hspace{0.5cm} \mbox{\sl ap}_1(L1,L2,L3).\\
\hspace{0.5cm} \mbox{\sl ap}_1(V,U,W),  & \mbox{\sl ap}_2([],L,L).\\
\hspace{0.5cm}\mbox{\sl perm}(W,T). & \mbox{\sl ap}_2([H|L1],L2,[H|L3])\leftarrow\\
& \hspace{0.5cm}\mbox{\sl ap}_2(L1,L2,L3).
\end{array}
\]

This example is analysed, based on Theorem~\ref{term-acc:term} for the 
well-moded case. We would like to prove termination of the goals 
$\mbox{\sl perm}(t_1, t_2)$, where $t_1$ is a ground list and $t_2$ a free 
variable.

Assume the modes 
\[\mbox{\sl perm}(\mbox{\sl in},\mbox{\sl out}), \mbox{\sl ap}_1(\mbox{\sl in},\mbox{\sl in},\mbox{\sl out}), \mbox{\sl ap}_2(\mbox{\sl out},\mbox{\sl out},\mbox{\sl in})\] 
The term-acceptability imposes, among the others, the following $>$-decrease:
$I\models \mbox{\sl ap}_2(V,[H|U],L)\theta\wedge\mbox{\sl ap}_1(V,U,W)\theta$ 
implies $\mbox{\sl perm}(L)\theta > \mbox{\sl perm}(W)\theta$. Note that the 
underlying decrease $L\theta > W\theta$ cannot be achieved by reasoning on 
$\mbox{\sl ap}_1/3$ or $\mbox{\sl ap}_2/3$ alone. Therefore,
we construct a following program $P'$:
\begin{eqnarray*}
&&\mbox{\sl ap}_2([],[t_1|t_2],[t_1|t_2]), \mbox{\sl ap}_1([],t_2,t_2).\\
&&\mbox{\sl ap}_2([t_6|t_1],[t_5|t_2],[t_6|t_3]),\mbox{\sl ap}_1([t_6|t_1],t_2,[t_6|t_4])\leftarrow \\
&& \hspace{1.0cm}\mbox{\sl ap}_2(t_1,[t_5|t_2],t_3), \mbox{\sl ap}_1(t_1,t_2,t_4).
\end{eqnarray*}
Now, we need to very that 
$M = \{\mbox{\sl ap}_2(a_1,a_2,a_3),\mbox{\sl ap}_1(b_1,b_2,b_3)\mid a_3 > b_3\}$ satisfies $T_{P'}(M)\subseteq M$. Using the 2 clauses, this
is reduced to ``$[t_1|t_2] > t_2$'' and
``$t_3 > t_4$ implies $[t_6|t_3] > [t_6|t_4]$'', imposing monotonicity and
subterm properties on $>$. The proof is completed analogously to the previous 
example.
$\hfill\Box$\end{example}

\eat{
\begin{example}
\begin{eqnarray*}
&& \mbox{\sl put}(X,s(0),X).\\
&& \mbox{\sl put}(X,s(s(Y)),X+K) \leftarrow put(X,s(Y),K).\\
&&\\
&& \mbox{\sl d}(\mbox{\sl der}(\mbox{\sl power}(X,0)), 0).\\
&& \mbox{\sl d}(\mbox{\sl der}(\mbox{\sl power}(X,s(Y))), K) \leftarrow \mbox{\rm var}(X),\mbox{\sl put}(\mbox{\sl power}(X,Y), s(Y), K).\\
&& \mbox{\sl d}(\mbox{\sl der}(X+Y),DX+DY) \leftarrow  \mbox{\sl d}(\mbox{\sl der}(X),DX), 
\mbox{\sl d}(\mbox{\sl der}(Y),DY).\\
&& \mbox{\sl d}(\mbox{\sl der}(\mbox{\sl der}(X)),DDX) \leftarrow \mbox{\sl d}(\mbox{\sl der}(X),DX), 
\mbox{\sl d}(\mbox{\sl der}(DX), DDX).\\
\end{eqnarray*}
Our aim is to prove termination for all atomic goals in the set 
\[S = \{\mbox{\sl d}(t_1, t_2)\mid t_1\;\mbox{\rm is a derivative of a polynomial and}\;\;t_2\;\mbox{\rm is a free variable}\}\] The set of calls, 
${\mbox {\sl Call}}(P,S)$ is 
\[
S\;\bigcup\;\left\{
\begin{array}{cc}
\mbox{\sl put}(t_1, t_2, t_3) & 
\left|
\begin{array}{l}
t_1\;\mbox{\rm is a symbolic power of the variable,}\\
t_2\;\mbox{\rm is an integer in the successor notation,}\\
t_3\;\mbox{\rm is a free variable}
\end{array}
\right.
\end{array}
\right\}\]

The requirement of rigidity of $>$ on ${\mbox {\sl Call}}(P,S)$
imposes that the second argument of $d$ and the third argument of 
${\mbox {\sl put}}$ should be ignored.
\eat{
Rigid term-acceptability implies, that for every $\theta$,
$\mbox{\sl put}(X,s(s(Y)),X+K)\theta > put(X,s(Y),K)\theta$ holds. By
pseudo-rigidity of $>$ on ${\mbox {\sl Call}}(P,S)$ holds, that for any
term $t$ $\mbox{\sl put}(t_1,t_2,Z) =_> put(t_1,t_2,t)$. Thus, in order
to enforce the inequality $\chi^{\cal M}_{\mbox {\sl put}}(2) = 1$ should
hold. The similar argument shows, that $\chi^{\cal M}_{\mbox {\sl d}}(1) = 1$.
}
Now we construct the decrease that follow from the rigid term-acceptability
condition, as usually by dropping that argument positions that cannot
affect $>$.

\begin{eqnarray}
&& \mbox{\sl put}(X,s(s(Y)))\theta > put(X,s(Y))\theta\\
&& \mbox{\sl d}(\mbox{\sl der}(\mbox{\sl power}(X, s(Y))))\theta > \mbox{\sl put}(\mbox{\sl power}(X, Y), s(Y))\theta\\
&& \mbox{\sl d}(\mbox{\sl der}(X+Y))\theta > \mbox{\sl d}(\mbox{\sl der}(X))\theta\\
&& I\models \mbox{\sl d}(\mbox{\sl der}(X), DX)\;\;{\mbox {\rm implies}}\;\;\mbox{\sl d}(\mbox{\sl der}(X+Y))\theta > \mbox{\sl d}(\mbox{\sl der}(Y))\theta\\
&& \mbox{\sl d}(\mbox{\sl der}(\mbox{\sl der}(X))) > \mbox{\sl d}(\mbox{\sl der}(X))\\
&& I\models \mbox{\sl d}(\mbox{\sl der}(X), DX)\;\;{\mbox {\rm implies}}\;\;\mbox{\sl d}(\mbox{\sl der}(\mbox{\sl der}(X))) >
\mbox{\sl d}(\mbox{\sl der}(DX))
\end{eqnarray}

Some of these inequalities can be achieved by imposing monotonicity
and subterm properties on different functors. 
\eat{
For example, to achieve (3)
$\chi^{\cal S}_{\mbox {\sl s}}(1) = 1$, to achieve (5) and (6)
$\chi^{\cal S}_{\mbox {\sl +}}(1) = \chi^{\cal S}_{\mbox {\sl +}}(2) = 1$,
and to achieve (7) $\chi^{\cal S}_{\mbox {\sl der}}(1) = 1$.
}
However, note that (8) cannot be achieved by imposing constraints on $>$.
Thus, some restrictions on the underlying model $I$  should be applied. In
our case the constraint is that $I$ should be such that for all
${\mbox {\sl d}}(t_1, t_2)\in I$: $\mbox{\sl der}(t_1) > \mbox{\sl der}(t_2)$.
To impose this constraint, we define $S = \{\mbox{\sl d}(t_1, t_2)\| 
der(t_1) > der(t_2)\}$ and enforce $T_P(S)\subseteq S$. This inclusion
causes the following constraints to arise:

\begin{eqnarray}
&& \mbox{\sl der}(\mbox{\sl der}(\mbox{\sl power}(t,0))) > \mbox{\sl der}(0)\\
&& I\models \mbox{\rm var}(t_1)\wedge \mbox{\sl put}(\mbox{\sl power}(t_1,t_2), s(t_2), t_3)\;\mbox{\rm implies} \\ \nonumber
&&\hspace{0.5cm}\mbox{\sl der}(\mbox{\sl der}(\mbox{\sl power}(t_1,s(t_2)))) > \mbox{\sl der}(t_3) \\
&& I\models \mbox{\sl der}(\mbox{\sl der}(t_1))>\mbox{\sl der}(t_3)\;\wedge\;
\mbox{\sl der}(\mbox{\sl der}(t_2))>\mbox{\sl der}(t_4)\;\mbox{\rm implies}\\ 
\nonumber
&&\hspace{0.5cm}\mbox{\sl der}(\mbox{\sl der}(t_1 + t_2)) > \mbox{\sl der}(t_3 + t_4)\\
&& I\models \mbox{\sl der}(\mbox{\sl der}(t_1))>\mbox{\sl der}(t_2)\;\wedge\;
\mbox{\sl der}(\mbox{\sl der}(t_2))>\mbox{\sl der}(t_3)\;\mbox{\rm implies}\\
\nonumber
&&\hspace{0.5cm}\mbox{\sl der}(\mbox{\sl der}(\mbox{\sl der}(t_1))) > \mbox{\sl der}(t_3)
\end{eqnarray}

In order to achieve condition (12) monotonicity of {\sl der} is imposed.
Further reduction constructs the following constraints:
\begin{eqnarray}
&&\mbox{\sl der}(\mbox{\sl power}(t, s(0))) > \mbox{\sl power}(t, 0)\\
&&I\models\mbox{\sl der}(\mbox{\sl power}(t_1, s(s(t_2)))) > t_3\;\mbox{\rm implies}\\ \nonumber
&&\hspace{1.0cm}\mbox{\sl der}(\mbox{\sl power}(t_1, s(s(s(t_2))))) > t_1 + t_3
\end{eqnarray}

The system (3)-(14) is consistent with the values of the characteristic
functions stated so far. For example, the recursive path ordering,
such that $\mbox{\sl der}\succ +$ and functors $+, s, \mbox{\sl power},
\mbox{\sl der}$ are monotone in all argument positions. 
satisfies it.
$\hfill\Box$\end{example}
}

As a last example, we return to the motivating Example~\ref{example:rep:der},
on computing higher derivatives of polynomial functions in one variable.
\begin{example}
We are interested in proving termination of the queries that belong to
$S = \{\mbox{\sl d}(t_1, t_2)\;\mid\;t_1\;\mbox{\rm is a repeated derivative of a function in a variable $u$ and}$ $t_2\;\mbox{\rm is a free variable}\}$. 
So $S$ consists of atoms of the form $\mbox{\sl d}(\mbox{\sl der}(u), X)$
or $\mbox{\sl d}(\mbox{\sl der}(u * u + u), Y)$ or 
$\mbox{\sl d}(\mbox{\sl der}(\mbox{\sl der}(u + u)), Z)$, etc.
Observe, that
${\mbox {\sl Call}}(P,S)$ coincides with $S$.

We start by analysing the requirements that imposes the rigidity of $>$ on
${\mbox {\sl Call}}(P,S)$. First, the second argument position of $d$ 
should be ignored, since it might be occupied by a free variable. Second, 
the first argument position of $d$ is occupied by a ground term. Thus, rigidity
does not pose any restrictions on functors argument positions.

Then, we construct the $>$-decreases that follow from the rigid 
term-acceptability.
The arguments that should be ignored are replaced by a term $t$. 
\begin{eqnarray}
&& \mbox{\sl d}(\mbox{\sl der}(X+Y)\theta , t) > \mbox{\sl d}(\mbox{\sl der}(X)\theta, t)\\
&& \mbox{\sl d}(\mbox{\sl der}(X),DX)\theta\;\;\mbox{\rm satisfies}\;\;R_{\mbox{\sl d}} \nonumber \\
&& \hspace{1.0cm}\mbox{\sl d}(\mbox{\sl der}(X+Y)\theta, t) > \mbox{\sl d}(\mbox{\sl der}(Y)\theta, t)\\
&& \mbox{\sl d}(\mbox{\sl der}(X*Y)\theta, t) > \mbox{\sl d}(\mbox{\sl der}(X)\theta, t)\\
&& \mbox{\sl d}(\mbox{\sl der}(X),DX)\theta\;\;\mbox{\rm satisfies}\;\;R_{\mbox{\sl d}} \nonumber \\
&& \hspace{1.0cm}\mbox{\sl d}(\mbox{\sl der}(X*Y)\theta, t) > \mbox{\sl d}(\mbox{\sl der}(Y)\theta, t)\\
&& \mbox{\sl d}(\mbox{\sl der}(\mbox{\sl der}(X))\theta, t) > \mbox{\sl d}(\mbox{\sl der}(X)\theta, t)\\
&& \mbox{\sl d}(\mbox{\sl der}(X),DX)\theta\;\;\mbox{\rm satisfies}\;\;R_{\mbox{\sl d}} \nonumber \\
&& \hspace{1.0cm}\mbox{\sl d}(\mbox{\sl der}(\mbox{\sl der}(X))\theta, t) > \mbox{\sl d}(\mbox{\sl der}(DX)\theta, t)
\end{eqnarray} 

Conditions (3)-(7) impose monotonicity and subset properties to hold on 
the first argument of $\mbox{\sl d}$.
In order to satisfy condition (8), it is sufficient to prove that 
for any $(t_1, t_2)\in R_{\mbox{\sl d}}$ holds that $t_1 > t_2$.
That is if $M = \{\mbox{\sl d}(t_1, t_2)\mid t_1 > t_2\}$ then 
$T_P(M)\subseteq M$. This may be reduced to the following conditions:
\begin{eqnarray}
&& \mbox{\sl der}(t) > 1\\ 
&& t_1\in R_{\mbox{\sl number}}\;\;\mbox{\rm implies}\;\;\mbox{\sl der}(t_1) > 0 \\
&& \mbox{\sl der}(t_1) > t_2\;\;\&\;\;\mbox{\sl der}(t_3) > t_4\;\;\mbox{\rm implies}\;\;\mbox{\sl der}(t_1 + t_3) > t_2 + t_4 \\
&& \mbox{\sl der}(t_1) > t_2\;\&\;\;\mbox{\sl der}(t_3) > t_4\;\;\mbox{\rm implies}\;\;\mbox{\sl der}(t_1 * t_3) > t_1 * t_4 + t_2 * t_3 \\
&& \mbox{\sl der}(t_1) > t_2\;\;\&\;\;\mbox{\sl der}(t_2) > t_3\;\;\mbox{\rm implies}\;\;\mbox{\sl der}(\mbox{\sl der}(t_1)) > t_3 
\end{eqnarray}
Condition (13) follows from monotonicity and transitivity of $>$. However,
(10)-(12) are not satisfied by general properties of $>$ and we choose, to
specify the order. The order that meets these conditions is the recursive 
path ordering~\cite{Dershowitz:RTA} with {\sl der} having the highest priority.
$\hfill\Box$\end{example}

This example demonstrates the main steps of our algorithm. First, given
a program $P$ and a set $S$ of goals, {\em compute the set of calls\/}
${\mbox {\sl Call}}(P,S)$ (for instance through
the abstract interpretation of~\cite{Janssens:Bruynooghe}). 
Second, {\em enforce the rigidity of $>$ on 
${\mbox {\sl Call}}(P,S)$}, i.e., ignore all predicate or functor
argument positions that might be occupied by free variables in 
${\mbox {\sl Call}}(P,S)$. Third, repeatedly {\em construct $>$-decreases},
such that rigid term-acceptability condition will hold and check if those
can be verified by some of the predefined orders. 

\section{Conclusion}
We have presented a non-transformational approach to termination analysis
of logic programs, based on general term-orderings. The problem of termination
was studied by a number of authors (see~\cite{DeSchreye:Decorte:NeverEndingStory} for the survey). More recent work on this topic can be found 
in~\cite{Lindenstrauss:Sagiv,Decorte:DeSchreye:98,Decorte:DeSchreye:Vandecasteele}.

Our approach gets it power from integrating the traditional 
notion of acceptability~\cite{Apt:Pedreschi:Studies} with the
wide class of term-orderings that have been studied in the context of the
term-rewriting systems. In theory, such an integration is unnecessary: acceptability (based on level mappings only) is already equivalent to LD-termination.
In practice, the required level mappings may sometimes be very complex (such
as for Example~\ref{example:rep:der} or {\sl distributive law}~\cite{Dershowitz:Hoot}, {\sl boolean ring}~\cite{Hsiang} or {\sl flattening of a binary tree}~\cite{Arts:PhD}), and automatic  systems for 
proving termination are unable to generate them. In 
such cases, generating an appropriate term-ordering, replacing the level
mapping, may often be much easier, especially since we can reuse the impressive
machinery on term-orders developed for term-rewrite systems. In some other
cases, such as {\sl turn}~\cite{Bossi:Cocco:Fabris}, simple level 
mappings do exist (in the case of {\sl turn}: a norm counting the number of 0s 
before the first occurrence of 1 in the list is sufficient), but most 
systems based on
level mappings will not even find this level mapping, because they only 
consider mappings based on term-size or list-length norms. Again, our approach 
is able to deal with such cases.

Unlike transformational approaches, that establish the termination results
for logic programs by the reasoning on termination
of term-rewriting systems, we apply the term-orderings directly to the logic
programs, thus, avoiding transformations. This could both be regarded as an 
advantage and as a drawback of our approach. It may be considered as a 
drawback, because reasoning on successful instances of intermediate body-atoms
introduces an additional complication in our approach, for which there is no
counterpart in transformational methods (except for in the transformation step
itself). On the other hand, we consider it as an advantage, because it is
precisely this reasoning on intermediate body atoms that gives more insight 
in the property of {\em logic program termination\/} 
(as opposed to {\em term-rewrite system
termination}). 

So, in a sense our approach provides the best of both worlds: a
means to incorporate into `direct' approaches the generality of general
term-orderings.

We consider as a future work a full implementation of the approach.
Although we already tested very many examples manually, an implementation 
will allow us to conduct a much more extensive experimentation, comparing
the technique also in terms of efficiency with other systems. Since we apply
a demand-driven approach, systematically reducing required conditions to more
simple constraints on the ordering and the model, we expect that the method can
lead to very efficient verification.

\section{Acknowledgement}
Alexander Serebrenik is supported by GOA: ``${LP}^{+}$: a second generation
logic programming language''. Danny De Schreye is a senior research associate
of FWO Flanders, Belgium.
 
\bibliography{/home/alexande/M.Sc.Thesis/main}
\bibliographystyle{abbrv}
\end{document}